\documentclass[pdftex,cite,aps,pra,amsmath,amsfonts,amssymb,superscriptaddress,twocolumn]{revtex4}
\renewcommand{\vec}[1]{\mathbf{#1}}
\usepackage{graphics}
\usepackage{braket}
\usepackage{wasysym}
\usepackage{setspace}

\usepackage[pdftex]{graphicx}

\renewcommand{\Im}{\operatorname{Im}}
\newcommand{\tr}{\operatorname{tr}}

\newcommand{\dS}{d\vec{S}}

\newcommand{\figref}[1]{Fig.~\ref{fig:#1}}

\renewcommand{\eqref}[1]{Eq.~(\ref{eq:#1})}
\newcommand{\Eqref}[1]{Equation~(\ref{eq:#1})}
\newcommand{\citeasnoun}[1]{Ref.~\onlinecite{#1}}
\newcommand{\secref}[1]{Sec.~\ref{sec:#1}}

\begin{document}

\title{Virtual photons in imaginary time: \\Computing exact Casimir forces via standard numerical-electromagnetism techniques}

\author{Alejandro Rodriguez}
\affiliation{Center for Materials Science and Engineering,
Massachusetts Institute of Technology, Cambridge, MA 02139}
\author{Mihai Ibanescu}
\affiliation{Center for Materials Science and Engineering,
Massachusetts Institute of Technology, Cambridge, MA 02139}
\author{Davide Iannuzzi}
\affiliation{Faculty of Sciences, Department of Physics and Astronomy, Vrije Universiteit Amsterdam, The Netherlands}
\author{J. D. Joannopoulos}
\affiliation{Center for Materials Science and Engineering,
Massachusetts Institute of Technology, Cambridge, MA 02139}
\author{Steven G. Johnson}
\affiliation{Center for Materials Science and Engineering,
Massachusetts Institute of Technology, Cambridge, MA 02139}


\begin{abstract}

We describe a numerical method to compute Casimir forces in arbitrary
geometries, for arbitrary dielectric and metallic materials, with
arbitrary accuracy (given sufficient computational resources).  Our
approach, based on well-established integration of the mean stress
tensor evaluated via the fluctuation-dissipation theorem, is designed
to directly exploit fast methods developed for classical computational
electromagnetism, since it only involves repeated evaluation of the
Green's function for imaginary frequencies (equivalently, real
frequencies in imaginary time).  We develop the approach by
systematically examining various formulations of Casimir forces from
the previous decades and evaluating them according to their
suitability for numerical computation.  We illustrate our approach with
a simple finite-difference frequency-domain implementation, test it
for known geometries such as a cylinder and a plate, and apply it to
new geometries.  In particular, we show that a piston-like geometry of
two squares sliding between metal walls, in both two and three
dimensions with both perfect and realistic metallic materials,
exhibits a surprising non-monotonic ``lateral'' force from the walls.
\end{abstract}


\maketitle

\section{Introduction}
\label{sec:intro}

One of the most dramatic manifestations of quantum mechanics observed
in the last half-century is the Casimir force: a tiny force on an
uncharged, source-free body due to changes in the zero-point energy
associated with quantum vacuum fluctuations (virtual
photons)~\cite{casimir, milonni,Lifshitz80, Plunien86,
Mostepanenko97}.  There have been many experimental verifications of
the Casimir force reported in recent decades~\cite{Lamoreaux97,
Onofrio06,Capasso07:review}, but always restricted to simple
geometries (parallel plates~\cite{bressi}, spheres and
plates~\cite{ninham,isra,over,lamo,moh1,moh2,moh3,hochan1,hochan2,decca1,decca2,pnas,lisanti,bressi},
or crossed cylinders~\cite{ninham,isra,ederth}).  Moreover, the force
in these particular geometries is almost always
attractive~\cite{KennethKl06} (except possibly for some unusual
material
systems~\cite{Boyer74,Hushwater96,Kenneth02,Iannuzzi03,ShaoZh06,Intravaia05,Sherkunov05,Fuchs07,Wetz01:arxiv,Capasso07:review})
and monotonically decreasing with separation. Thus, one might ask
whether it is possible to obtain non-monotonic or even repulsive
forces in more complex structures, and more generally whether complex
geometries might give rise to unexpected force phenomena. For more
complicated geometries, however, calculations become extremely
cumbersome and often require drastic approximations, a limitation that
has hampered experimental and theoretical work beyond the standard
geometries.

In this paper, we explore several ways in which well-established,
efficient techniques from standard computational electromagnetism can
be brought to bear on this problem, in order to predict forces for
arbitrary geometries and materials with arbitrary accuracy (no
uncontrolled approximations).  Starting from the simplest, most direct
approaches, we show that practical considerations naturally lead
towards a particular method involving the integral of the Minkowski
stress tensor by repeated evaluation of the imaginary-frequency
Green's function---a method previously developed for purely analytical
calculations~\cite{lifshitz1,Dzyaloshinskii61,Lifshitz80}.  We
illustrate the method by a simple finite-difference implementation,
but evaluation of the Green's function is so standard that many more
sophisticated techniques are immediately applicable, and we discuss
what techniques are likely to be optimal.  Our approach is tested for
geometries with known solutions, and then is applied to new geometries
in two and three dimensions that lead to surprising non-monotonic
effects.  We also demonstrate the application of our technique to
dispersive dielectric materials, not just for idealized perfect
metals.  We explain how the same technique can be used for
visualization of the Casimir interactions between bodies, as well as
for computing other quantities of interest, such as torques.  The key
advantage of exploiting standard computational approaches is not
merely that existing code, error analyses, and other experience can be
applied to the Casimir problem, but also that these methods have been
proven to scale to large three-dimensional problems, which have
previously seemed out of reach of exact methods for the Casimir force.

The most common approach to predicting the Casimir force has been to
consider approximations for small perturbations around known
solutions, such as parallel plates or dilute gases.  For parallel
plates in $d$ dimensions, separated by a distance $a$, there is a
well-known attractive force that scales as $1/a^{d+1}$, first
predicted by Casimir~\cite{casimir} and later extended to formulas for
any planar-multilayer dielectric distribution $\varepsilon(x,\omega)$
via the generalized Lifshitz formula~\cite{Tomas02}.  A direct,
intuitive extension of this result is the proximity force
approximation (PFA)~\cite{bordag01}, which treats the force between
two surfaces as a two-body interaction given by the sum of ``parallel
plate'' contributions.  Valid in the limit of small curvature, PFA
provides an easy way to conceptualize the Casimir force in complex
geometries as a simple two-body force law, but unfortunately it may
also be deceptive: outside its range of applicability, the Casimir
force is not additive~\cite{milonni} and may be qualitatively
different from PFA's
predictions~\cite{emig01,genet03,emig03_1,emig03_2,gies06:PFA,maianeto05,Rodriguez07:PRL}.
Other perturbative approaches include renormalized
Casimir-Polder~\cite{Tajmar04, Sedmik06} or semi-classical
interactions~\cite{Schaden98}, multiple scattering
expansions~\cite{Balian78, Lambrecht06}, classical ray optics
approximations~\cite{Jaffe04}, higher-order PFA
corrections~\cite{Bordag06}, and other perturbative
techniques~\cite{Emig03}.  The ray-optics approach is especially
interesting because, although it is only strictly valid in the
small-curvature limit, it captures multiple-body interactions and can
therefore sometimes predict the qualitative behavior in cases where
other approximations
fail~\cite{Jaffe04:preprint,Brevik05:pfa,Jaffe07:ray}.  Nevertheless,
none of these methods provide any guarantees of accuracy in arbitrary
geometries, where they involve uncontrolled approximations.
Therefore, for complex new geometries, where one might hope to
encounter behaviors very different from those in the parallel-plate
limit, a different approach is required.

To this end, researchers have sought ``exact'' numerical methods
applicable to arbitrary geometries---that is, methods that converge to
the exact result with arbitrary accuracy given sufficient
computational resources. One such method was proposed by
\citeasnoun{emig05}, based on a path-integral representation for the
effective action; this method has predicted the force between a
cylinder and a plate~\cite{emig06}, and between corrugated
surfaces~\cite{emig03_1,emig03_2}.  It is based on a surface
parameterization of the fields coupled via vacuum Green's functions,
requiring $O(N^2)$ storage and $O(N^3)$ time for $N$ degrees of
freedom, making scaling to three dimensions problematic. Another exact
method is the ``world-line approach''~\cite{gies03}, based on
Monte-Carlo path-integral calculations. The scaling of the world-line
method involves a statistical analysis, determined by the relative
feature sizes in the geometry, and is discussed below.  The methods of
\citeasnoun{emig05} and \citeasnoun{gies03} have currently only been
demonstrated for perfect-metallic $z$-invariant structures---in this
case, the vector unknowns can be decomposed into TE
($\vec{E}\cdot\hat{\vec{z}}=0$) and TM ($\vec{H}\cdot\hat{\vec{z}}=0$)
scalar fields with Neumann and Dirichlet boundary conditions,
respectively---although generalizations have been
proposed~\cite{gies03, emig04_1}. Here, we propose a method based on
evaluation of the mean Minkowski stress tensor via the
fluctuation-dissipation theorem, which only involves repeated
evaluation of the electromagnetic imaginary-frequency Green's
function.  For our initial volume discretization with $N$ degrees of
freedom and an efficient iterative solver, this requires $O(N)$
storage and at best $O(N^{2-1/d})$ time in $d$
dimensions. Furthermore, because evaluation of the Green's function is
such a standard problem in classical computational electromagnetism,
it will be possible to exploit many developments in fast solvers,
based on finite-element~\cite{chew01,Volakis01,Jin02,Zhu06},
spectral~\cite{boyd01:book,chew01}, or boundary-element
methods~\cite{Hackbush89,chew97,Rao99,chew01, Volakis01,Jin02}.  As we
argue below, a future implementation using boundary-element methods
should attain nearly $O(N \log N)$ time.  To illustrate the method,
however, our initial implementation is based on the much simpler
finite-difference frequency-domain method~\cite{Christ87} with a
conjugate-gradient solver~\cite{bai00}, as described below.

In the following sections, we describe the step-by-step conceptual
development of our computational method.  Our purpose here is to start
back at the beginning, with the earliest theoretical descriptions of
the Casimir force, and analyze these formulations from the point of
view of their suitability for purely numerical calculations.  Although
the final technique, in terms of the stress tensor integrated over
space and imaginary frequency, can be viewed as a numerical
implementation of a textbook result due to Dzyaloshinski\u{\i}
\textit{et al.}~\cite{Dzyaloshinskii61,Lifshitz80,Pitaevski06}, it is
illustrative to derive it as the culmination of a sequence of simpler
approaches, in order to show how it circumvents a number of numerical
obstacles that would hinder a more direct method.  To begin with, we
illustrate the methods using the well-known case of parallel plates
where they can be compared to analytical expressions, but a more
rigorous test is subsequently provided by the situation of a cylinder
and plate, recently solved numerically~\cite{emig06}.  Finally, we
apply our method to a new geometry of a ``piston''-like structure
involving blocks sliding between parallel walls, in both two and three
dimensions with both perfect metals and more realistic dispersive
dielectrics, and demonstrate a surprising non-monotonic ``lateral''
effect from the walls.  We conclude by analyzing the scaling of the
method compared to previous approaches and discussing the application
of more sophisticated finite-element and boundary-element techniques.

\section{A Simplistic Approach}
\label{sec:simplistic}

Perhaps the simplest approach to computing the Casimir force is to think
of it as the derivative of the zero-point energy $U$ expressed as a
sum of ground-state photon energies $\hbar\omega/2$.  For each photon
mode with frequency $\omega$, the zero-point energy is $\hbar \omega /
2$, and thus the total Casimir energy, at least for non-dissipative
systems where $\omega$ is real~\cite{Tomas02}, is formally given by
the sum over all modes~\cite{casimir,milonni}:
\begin{equation}
  U = \sum_{\omega} \frac{1}{2} \hbar \omega
\label{eq:U}
\end{equation}
This sum is formally infinite, because the classical harmonic modes
$\omega$ have unbounded frequencies.  There is some controversy over
the physical interpretation of this divergence~\cite{milton04}, but in
practice it is removed by regularizing the sum in some fashion, for
example multiplying by $e^{-s\omega}$ for $s>0$, and taking
$s\rightarrow 0$ only \emph{after} the sum is differentiated to obtain
the force $F = -dU/da$ between two bodies with separation
$a$~\cite{Mazzitelli06}.  This approach, which was an early method to
analytically compute the force between perfect-metal
plates~\cite{casimir}, might seem to provide the most direct
computational method as well.  After all, the computation of
electromagnetic eigenmodes is routine even in very complicated
geometries, and efficient methods are known~\cite{Johnson2001:mpb,
chew01, Volakis01, Jin02}.  Unfortunately, it turns out not to be
practical for this problem (except in one-dimensional
geometries~\cite{Enk95}), as explained below, but the \emph{reason}
why it is impractical points the way to more efficient methods.

To illustrate the difficulty in directly evaluating \eqref{U}, let us
consider the simplest one-dimensional geometry: two parallel
perfect-metal plates, separated by a distance $a$, in which case one
can predict analytically an attractive force $F = \pi\hbar
c/24a^2$~\cite{milton04}.  Ignoring this analytical result, let us
apply a numerical method that, conceptually, we could apply to an
arbitrary geometry:
\begin{enumerate}
\item Discretize space with resolution $\Delta x$ using a
finite-difference approximation, with space truncated to a finite
computational cell (e.g. with periodic boundaries).
\item Solve numerically for the eigenmode frequencies $\omega$ and sum to obtain $U(a)$.
\item Shift one body (one plate) by one pixel $\Delta x$ and thus
compute $U(a+\Delta x)$.
\item Obtain the force $F \approx -[U(a+\Delta x) - U(a)] / \Delta x$.
\end{enumerate}
Note that this method automatically provides its own regularization:
the number of modes $\omega$ in a discretized computational cell is
finite (the frequencies are bounded by the Nyquist limit), and hence
$U$ is finite for $\Delta x > 0$.  The periodic boundaries will lead
to artificial ``wrap-around'' forces, but since Casimir forces decay
rapidly with distance, the contribution of these forces can be made
negligible for a sufficiently large computational cell.

\begin{figure}[hbt]
\includegraphics[width=0.45\textwidth]{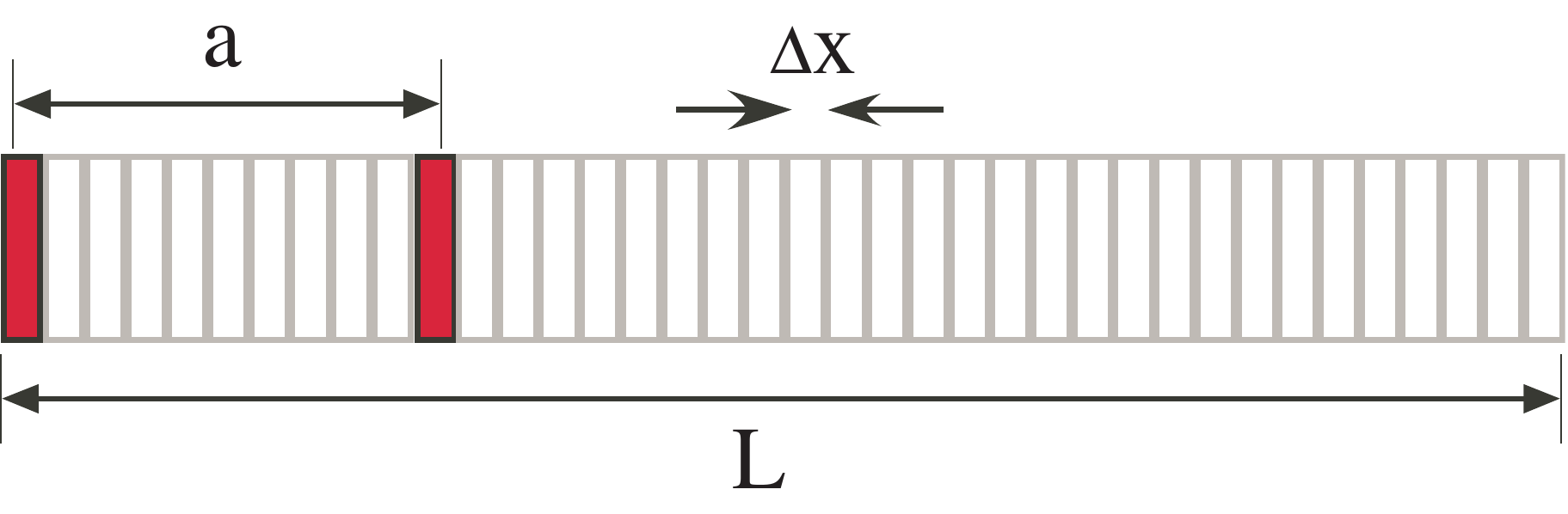}
\caption{(Color) Schematic of one dimensional geometry, showing two 1d
  metal plates separated by a distance $a$, embedded in a
  computational cell of length $L=a+4a$, with periodic boundary
  conditions, and resolution $\Delta x$.}
\label{fig:1d-model}
\end{figure}

This method, for the one-dimensional parallel-plate geometry, is
illustrated in \figref{1d-model}. Here, we have two plates with
separation $a$ and an overall computational cell size $L = 5a$ (which
will contribute an erroneous wrap-around force 1/16 of the physical
force).  Maxwell's equations, in one dimension, can be written as the
scalar eigenproblem $\nabla^2 E_z = \omega^2 E_z$ (in $c=1$ units),
which is discretized to
\begin{equation}
\frac{E_{n+1} - 2E_n + E_{n-1}}{\Delta x^2} = \omega_n^2 E_n
\end{equation}
in a center-difference approximation for $E_z(n\Delta x) = E_n$.  For two metal plates with separation $d$ and Dirichlet boundary condition $E_z = 0$, the discrete eigenvalues $\omega_n$ can be found analytically:
\begin{equation}
\omega_n = \frac{2}{\Delta x} \sin\left(\frac{n\pi\Delta x}{2d}\right),
\end{equation}
for $n = 0, \ldots, d/\Delta x$.  The energy $U$ is then given by
summing $\omega_n$ in \eqref{U} for $d=a$ and $d = L-a$, and the force
$F$ by the discrete derivative of $U$ as above.

Applying this procedure numerically for $\Delta x = 0.05a$, one
obtains the correct force to within 5\% (and to any desired accuracy
by increasing $L$ and decreasing $\Delta x$), so at first glance it
may seem that the method is successful.  However, its impracticality
is revealed if we examine the contribution of each frequency
$\omega_n$ to the force.

\begin{figure}[hbt]
\centerline{\includegraphics[width=0.5\textwidth]{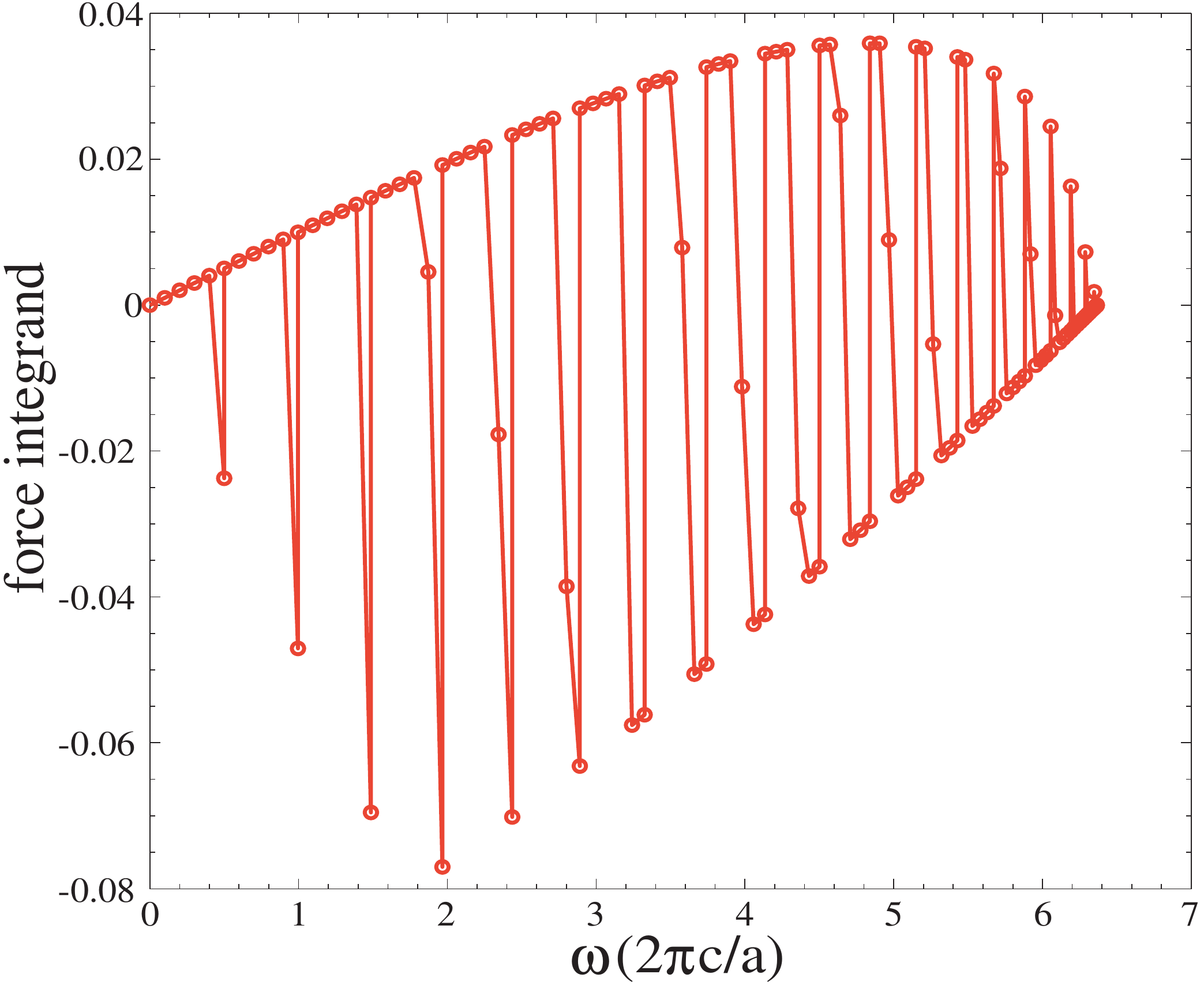}}
\centerline{\includegraphics[width=0.5\textwidth]{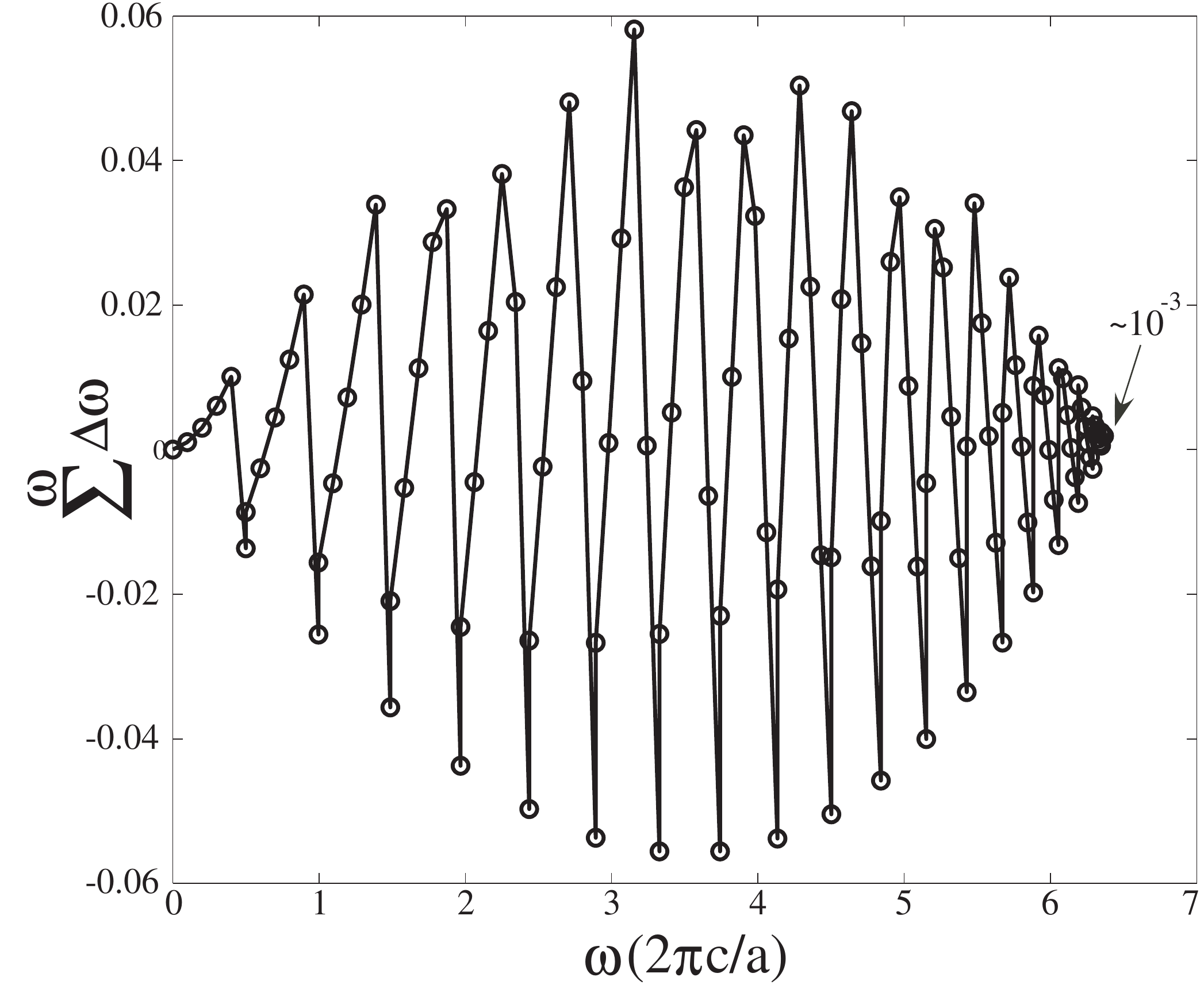}}
\caption{(Color) \emph{top}: Plot of force summand, or spectral density,
$\hbar \Delta \omega / 2\Delta x$ vs. $\omega$ for 1d parallel metal plates from \figref{1d-model}. \emph{bottom}: Plot of force partial sum
$\sum^{\omega} \hbar \Delta \omega / 2\Delta x$ vs. $\omega$.}
\label{fig:fdensity}
\end{figure}

The top panel of \figref{fdensity} shows the contribution $\hbar
\Delta\omega/2\Delta x$ [$\Delta\omega = \omega(a+\Delta x) -
\omega(a)$] to the force summation, and the bottom panel shows the
corresponding partial sum (for frequencies up to $\omega$).  We
see that \emph{every} frequency (of the regularized/finite-resolution
problem) makes a non-negligible contribution to the force, and the
summation is of a wildly oscillating quantity that leaves a tiny
remainder at the end.  The reason for these oscillations is quite
simple: as $a$ is increased, the frequencies on the $4a$ side of the
plates increase slowly, while the smaller number of frequencies on the
$a$ side of the plate decrease more rapidly, and these lead to the
positive and negative contributions in the top panel of
\figref{fdensity}, respectively.

These two features, which are intrinsic properties not limited to this
particular discretized geometry~\cite{Ford93}, combine to make this
method impractical in higher-dimensional structures.  Because
\emph{every} frequency contributes to the force, in a numerical method
one must compute every eigenvalue of the Maxwell eigenproblem.  In one
dimension, that is not so bad, but in general if there are $N$ degrees
of freedom ($N$ grid points), then computing every eigenvalue of an $N
\times N$ matrix requires $O(N^2)$ storage and $O(N^3)$ time.  This is
impractical in three dimensions where $N$ may be in the millions.
Furthermore, the wild oscillations of the summand imply that the
eigenvalues must be computed quite accurately, and may exacerbate
numerical difficulties in larger problems.

However, these undesirable features are avoidable, because we have not
yet exploited a key property of Maxwell's equations: causality.  If we
ignore the causality constraint, then the oscillatory spectrum would
be an observable effect: one would simply employ a material that is a
perfect metal in some frequency range and transparent otherwise, in
order to obtain the force spectrum integrated only in that range.  Such
a material, however, would violate the Kramers-Kronig constraints that
follow from causality considerations~\cite{Jackson98}.  Thus, we are
motivated to exploit causality in some fashion to avoid the
oscillatory spectrum.

\section{Wick Rotation and Energy Density}
\label{sec:energy}

In order to exploit causality, we will rewrite \eqref{U} in terms of
the electromagnetic Green's function via an integral over the density
of states.  Causality implies that the Green's function has no poles
in the upper-half plane, so one can perform a contour integration, or
\emph{Wick rotation}, to transform the sum over real frequencies into
an integral along the imaginary-frequency axis.  The result of this
standard trick turns out to be a well-known expression: an integral of
the mean electromagnetic energy density, evaluated by the
fluctuation-dissipation theorem using the temperature (Matsubara)
Green's function.  Again, we focus on the method's suitability as a
purely \emph{numerical} approach, for arbitrary geometries, and we
will find that it still leaves something to be desired.

First, we can express the zero-point energy of \eqref{U} as an
integral over the local density of states $D(\vec{x}, \omega)$:
\begin{equation}
U = \frac{\hbar}{2}\int_0^\infty d\omega \int \omega D(\vec{x},\omega) \, d^3\vec{x}.
\label{eq:energy-DOS}
\end{equation}
Since we are solving for the eigenstates of Maxwell's equations
$(\nabla \times \nabla \times -\omega^2\varepsilon)\vec{E} = 0$, the
local density of states $D(\vec{x},\omega)$ can be expressed in terms
of the Green's tensor $G_{jk}$~\cite{Landau:QM}:
\begin{widetext}
\begin{eqnarray}
  D(\vec{x},\omega) =\; \frac{1}{\pi}
 \frac{d(\omega^2\varepsilon)}{d\omega} \sum_{k=1}^3 \; \Im
 \braket{\vec{x;\hat{\vec{e}}_k} | \frac{1}{\nabla \times \nabla
 \times - \; \omega^2\varepsilon + i0^{+}} | \vec{x; \hat{\vec{e}}_k}}
 = \; \frac{1}{\pi} \frac{d(\omega^2\varepsilon)}{d\omega} \Im \tr
 G(\omega; \vec{x}-\vec{x})
\end{eqnarray}
\end{widetext}
where $G_{jk}$ solves: $(\nabla \times \nabla \times - \;
\omega^2\varepsilon) \vec{G}_k(\omega; \vec{x}-\vec{x'}) =
\delta^{3}(\vec{x}-\vec{x'}) \hat{\vec{e}}_k$, with $\hat{\vec{e}}_k$
denoting the unit vector in the $k$th direction.  For non-dissipative
systems in which \eqref{U} is valid, $\varepsilon$ is real and we can
therefore pull the $\Im$ outside of the integral.  (The generalization
to dissipative materials is discussed below.)

Furthermore, we know from causality requirements that the Green's
function has no poles in the upper half plane in
$\omega$-space~\cite{Jackson98, milonni}.  This means that one can
perform a contour integration to relate $\int_0^\infty d\omega$ to the
integral $\int_0^\infty dw$ along the imaginary-frequency axis
$\omega=iw$, also known as a Wick rotation.  We therefore obtain:

\begin{equation}
  U = \frac{\hbar}{2\pi} \int^{\infty}_0 dw \int w
  \frac{d[w^2\varepsilon(iw)]}{dw} \tr G(iw; \vec{x}-\vec{x}) \, d^3\vec{x},
\label{eq:Uimag}
\end{equation}
where the \emph{new} problem to be solved is that of finding the
solutions to the imaginary-time Green's function ($c=1$ units): 
\begin{equation}
  \left[\nabla \times \nabla \times + w^2\varepsilon(iw,\vec{x})\right]
  \vec{G}_k(iw; \vec{x}-\vec{x'}) = \delta^3(\vec{x}-\vec{x'}) \hat{\vec{e}}_k.
\label{eq:Green}
\end{equation}
As usual, this is formally infinite, because the Green's function is
singular at $\vec{x} = \vec{x}'$, but one typically regularizes the
problem by subtracting the vacuum Green's function, which removes the
singularity without changing the net force.

\Eqref{Uimag} is not a new result, nor is it limited to
non-dissipative materials (unlike our
derivation)~\cite{Lifshitz80,Tomas02}.  In fact, it is equivalent to
the mean energy in the fluctuating electromagnetic field, derived from
the fluctuation-dissipation theorem via the temperature Green's
functions~\cite{Lifshitz80}.  Our purpose in deriving it this way is
to emphasize the connection to the simplistic approach of \eqref{U}.
In particular, the mean energy in the electromagnetic fields (for the
case of non-magnetic materials $\mu = 1$) is given
by~\cite{Lifshitz80,Jackson98}:
\begin{eqnarray}
  U &=& \int_0^\infty dw \int \frac{1}{2} \left[ \frac{d(w \varepsilon)}{dw}
\left\langle \vec{E}^2 \right\rangle_w + \left\langle \vec{H}^2
\right\rangle_w\right] d^3\vec{x} \nonumber \\ &=& \int_0^\infty dw
\int \frac{1}{2w} \frac{d(w^2 \varepsilon)}{dw} \left\langle \vec{E}^2
\right\rangle_w d^3\vec{x},
\label{eq:Uenergy}
\end{eqnarray}
using the fact that $\int \varepsilon \langle \vec{E}^2 \rangle_w =
\int \langle \vec{H}^2 \rangle_w$.  Here, the key point is that the
mean values of the fluctuating fields are given, via the
fluctuation-dissipation theorem, in terms of the imaginary-frequency
Green's function~\cite{Lifshitz80}:
\begin{equation}
\left\langle E_{j}(\vec{x}) E_{k}(\vec{x'})\right\rangle_w 
    =\frac{\hbar}{\pi} w^2 G_{jk}(iw;\vec{x} - \vec{x'}).
\label{eq:Ecorr}
\end{equation}
where the dyadic Green's function $G_{ij}$ solves \eqref{Green} and
obeys the usual boundary conditions on the electric field from
classical electromagnetism~\cite{Jackson98}. Substituting
\eqref{Ecorr} into \eqref{Uenergy}, one recovers \eqref{Uimag}.

From a computational perspective, the imaginary-frequency integral of
\eqref{Uimag} turns out to be greatly superior to the real-frequency
summation of \eqref{U}, for two reasons.  First, while every real
frequency $\omega$ contributed to the force $-dU/da$, the same is not
true for the derivative of the imaginary-frequency integrand.  In
particular, as discussed below, the force integrand in imaginary
frequencies turns out to be a smooth, non-oscillatory, strongly peaked
function of $w$, meaning that one can integrate it via a
smooth-quadrature method that evaluates the integrand at only a small
number of $w$ values.  Second, the imaginary-frequency Green's
function turns out to be quite easy to obtain by relatively standard
methods, including for dissipative systems (where obtaining the
eigenmodes is harder because it involves a non-Hermitian
eigenproblem).  These two favorable features are closely related.

Consider \eqref{Green} for the imaginary-frequency Green's function.
Unlike its real-frequency counterpart, the linear operator on the
left-hand side of this equation is \emph{real-symmetric} and
\emph{positive-definite} (for $w>0$).  This is true even for
dissipative materials where $\varepsilon(\omega)$ is complex, since
causality requirements imply that $\varepsilon(iw)$ is purely real and
positive (in the absence of gain)~\cite{Jackson98, milonni}.  For one
thing, this implies that the most powerful numerical methods are
applicable to solving the linear system of \eqref{Green}---many of
these methods (e.g. the conjugate-gradient method) are restricted to
Hermitian positive-definite operators~\cite{bai00}.  Also, the
resulting Green's function is particularly well-behaved: it is
exponentially decaying and non-oscillatory.  This transforms the
highly oscillatory real-$\omega$ force integrand into a mostly
non-oscillatory integrand, and also makes the force integrand
exponentially decaying for large $w$ (for large $w$, the interactions
between bodies become exponentially small).  (In addition, as we will
discuss in \secref{beyond-FDFD}, the exponentially decaying Green's
function is especially favorable for boundary-element numerical
methods.)

Again, considering the simplest possible finite-difference scheme,
this leads us to the following numerical algorithm to compute the
force:
\begin{enumerate}

\item For a given imaginary frequency $w$:
\begin{enumerate}
\item For every grid point $\vec{x}$, solve \eqref{Green} for each polarization $k$ to obtain $G_{kk}(iw;\vec{x}-\vec{x})$.
\item Sum over $\vec{x}$ to compute the spatial integral in \eqref{Uimag}.
\item Repeat the above for the body shifted by one pixel $\Delta x$ and subtract to obtain the force integrand at $w$.
\end{enumerate}

\item Employ a smooth quadrature scheme to integrate the above
function over $w$ to obtain the force.

\end{enumerate}
Again, the spatial discretization provides its own regularization
($G_{kk}$ is finite), and thus no additional regularization is
required (the contribution of the vacuum Green's function to the net
force is zero).  Again, one can truncate the computational cell in a
number of ways, for example with periodic boundaries, and the
artifacts thereby introduced will decrease rapidly with cell size.
Again, there are also many other ways that one could potentially solve
for $G_{kk}$ besides a finite-difference approximation, but we will
delay discussion of those techniques until we have formulated the
final method in the next section.

\begin{figure}
\centerline{\includegraphics[width=0.5\textwidth]{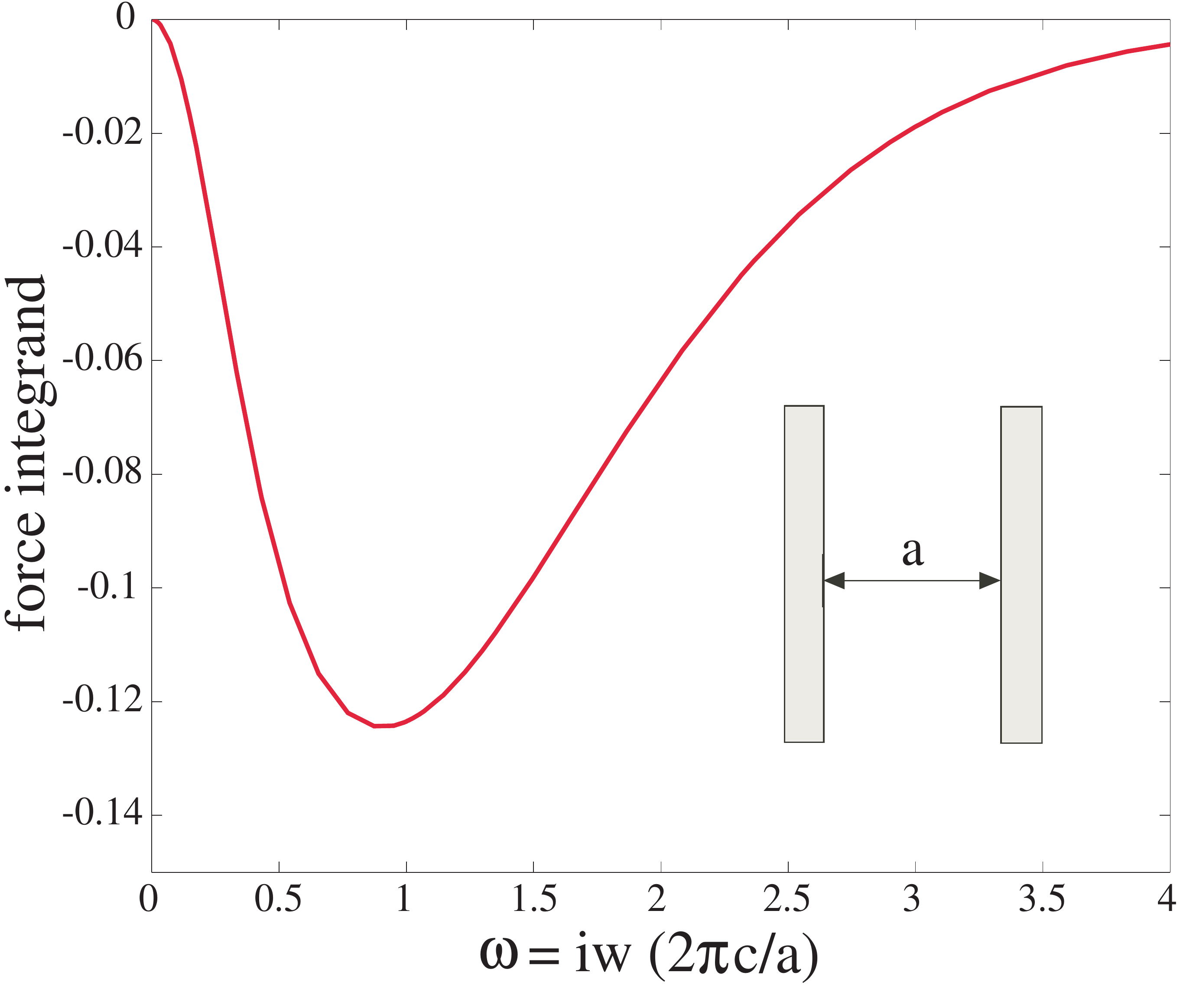}}
\caption{(Color) Plot of Casimir force integrand $dU/da$ between two 1d
parallel plates separated by a distance $a=1$ versus imaginary
frequency $w = \Im \omega$, using the method of \secref{energy}.}
\label{fig:psum-casimit}
\end{figure}

The above procedure can again be applied to the one-dimensional
problem of the force between two plates, as in \figref{fdensity}, to
illustrate its basic features.  The result is shown in
\figref{psum-casimit}, plotting the force integrand as a function of
imaginary frequency $w$, and the difference from \figref{fdensity} is
striking.  The integrated force is the same (correct) result as
before.  Unlike \figref{fdensity}, the force integrand has no sign
oscillations, is exponentially decaying for large $w$, and is strongly
peaked around a characteristic $w = 2\pi c/a$, corresponding to a
``wavelength'' of $a$ (the separation).  These features imply that the
force can be accurately integrated by an adaptive Gauss-Kronrod
quadrature scheme~\cite{piessens83} using at most a few dozen $w$
points.

Although this method is much more efficient than the one described in
the previous section, and is potentially practical at least in two
dimensions, it still has some undesirable features.  Suppose that we
have $N$ grid points in our discretized operator ($N$ may be very
large in 3d).  Even if we have an ideal iterative solver for the
sparse linear system of \eqref{Green}, such as an ideal multigrid
solver~\cite{Trottenberg01, Zhu06}, each evaluation of the Green's
function takes at best $O(N)$ time with $O(N)$ storage.  However, we
must evaluate the Green's function $3N$ times in order to perform the
spatial integration, resulting in $O(N^2)$ complexity.  As is
discussed in the next section, one can do much better than this by
using the \emph{stress tensor} instead of the energy density.  In
fact, as is discussed in \secref{beyond-FDFD}, it should ultimately be
possible to obtain the force with nearly $O(N \log N)$ work using a
boundary-element method to compute the stress tensor, in which $N$ is
only the number of degrees of freedom required to represent the
\emph{interfaces} between materials.

\section{Stress-tensor Computational Approach}
\label{sec:stress-approach}

After having analyzed the feasibility of several techniques to solve
the Casimir problem through the lens of numerical electromagnetism, we
are ready to appreciate and explore the most feasible of the methods
thus far presented: an approach based on the Maxwell stress
tensor. 

As derived by Dzyaloshinski\u{\i}~\textit{et
al.}~\cite{Dzyaloshinskii61,Lifshitz80,Pitaevski06}, the net Casimir
force on a body can be expressed as an integral over any closed
surface around the body of the mean electromagnetic stress tensor
$\langle T_{ij} \rangle$, integrated over all
frequencies. Again, using the same arguments as above, it is
computationally convenient to perform a Wick rotation, expressing the
net force as an integral over imaginary frequencies $\omega=iw$.  (The
original derivation used imaginary frequencies to start with, via the
temperature Green's function, but the result is equivalent to a Wick
rotation of the real-frequency expression.  A related analytical
treatment, but using purely real $\omega$ and therefore unsuitable for
numerical computation because of the oscillations discussed above, has
also been examined~\cite{MatloobKe99,Matloob99}.)  The resulting net
force is:
\begin{equation}
\vec{F} = \int_0^\infty dw \oiint_{\mathrm{surface}} \langle \mathbf{T}(\vec{r},iw) \rangle \cdot \dS \, .
\label{eq:F}
\end{equation}
In two or one dimensions, one or two of the spatial integrals are
omitted, respectively, but the result still has the units of force;
the change in dimensions of $\dS$ is balanced by a change in the
dimensions of the delta function in \eqref{Green}.  On the other hand,
for a 3d structure that is $z$-invariant [a constant 2d cross-section
$\varepsilon(x,y)$] or $yz$-invariant [a univariate $\varepsilon(x)$],
the integrals over the invariant directions are replaced by integrals
over the corresponding wavevectors, resulting in a net force per unit
length or per unit area, respectively. (These wavevector integrals are
discussed in more detail in \secref{periodic-forces}.)  The stress
tensor is defined as usual by:
\begin{widetext}
\begin{align}
\left\langle T_{ij} (\vec{r},iw) \right\rangle = \mu(\vec{r},iw) \left[ \left\langle
H_{i}(\vec{r})\,H_{j}(\vec{r})\right\rangle
-\frac{1}{2}\delta_{ij}\sum_k\left\langle
H_{k}(\vec{r})\,H_{k}(\vec{r})\right\rangle \right]  +
\varepsilon(\vec{r},iw)\left[\left\langle
E_{i}(\vec{r})\,E_{j}(\vec{r})\right\rangle -\frac{1}{2}\delta_{ij}\sum_k
\left\langle E_k(\vec{r})\,E_k(\vec{r})\right\rangle \right] 
\label{eq:stress}
\end{align}
\end{widetext}
where $\mu$ and $\varepsilon$ are the relative permeability and
permittivity, respectively, although in most cases we set $\mu=1$ for
simplicity (since most materials have negligible magnetic response at
short wavelengths, and in any case the stress tensor is normally
evaluated over a surface lying in vacuum).  As before, the connection
to quantum mechanics arises from the correlation functions of the
fluctuating fields, given via the fluctuation-dissipation theorem in
terms of the imaginary-$\omega$ dyadic Green's function. The
correlation function for the electric field $\langle E_iE_j \rangle$
is given in \eqref{Ecorr}. In this case, however, we also need the
magnetic-field correlation functions, which can be obtained by
differentiating the electric-field Green's function~\cite{Lifshitz80}:
\begin{equation}
\left\langle
    H_{i}(\vec{r}) H_{j}(\vec{r}')\right\rangle
    =-\frac{\hbar}{\pi} (\nabla\times)_{i\ell}(\nabla'\times)_{jm} G_{\ell m}(iw;\vec{r}
    - \vec{r}') \, ,
\label{eq:Hcorr}
\end{equation}
(Alternatively, the $\langle H_i H_j \rangle$ correlation function can
be computed from the magnetic Green's function, which is the magnetic
field in response to a given magnetic-dipole current.)  The above
expressions are given at zero temperature; the nonzero-temperature
force is found by changing $\int dw$ in \eqref{F} into a discrete
summation~\cite{Lifshitz80,Lambrecht06}.  Although the Green's
function (and thus $\mathbf{T}$) is formally infinite at $\vec{r} =
\vec{r}'$, this divergence is conventionally removed by subtracting
the vacuum Green's function; in a numerical method with discretized
space, as below, there is no divergence and no additional
regularization is required.  (The vacuum Green's function gives zero
net contribution to the $\dS$ integral, and therefore need not be
removed as long as the integrand is finite.)

Historically, this stress-tensor expression was used to derive the
standard Lifshitz formula for parallel plates, where $G_{ij}$ is known
analytically. Its adaptability and suitability as a purely
computational method does not seem to have been recognized, however.
As in the previous section, the method involves computing the Green's
function for many imaginary frequencies $w$ and spatial points
$\vec{x}$, integrated over $w$ and $\vec{x}$. However, a quick glance
at \eqref{F} will suggest at least two obvious computational
advantages compared to the method discussed in \secref{energy}. First,
in framing the problem in terms of the stress tensor, we have reduced
the spatial integral over the whole volume (\eqref{Uenergy}) to a
surface integral around the body of interest. This implies that, for a
$d$-dimensional geometry, the computational effort due to spatial
integration is reduced from $O(N^2)$ to $O(N^{2-1/d})$. Second, the
force is now given \emph{directly} in terms of the dyadic Green's
function (via the stress tensor), rather than its derivative, which
avoids another layer of computation. Moreover, although our derivation
is only valid when the stress tensor is evaluated at points within
lossless dielectrics (regardless of whether the bodies themselves are
dissipative), one can also extend it for evaluation in absorbing
media~\cite{Pitaevski06}. However, the case discussed above (bodies
separated by vacuum) is the most common.

So far, we have presented the step-by-step development of an efficient
approach to computing Casimir forces. In what follows, we illustrate
our new approach using a proof-of-concept finite-difference
implementation, and present results that demonstrate its flexibility
and utility.

\section{The Finite-Difference Method}
\label{sec:FDFD}

At this point, all that remains is the numerical computation of the
Green's function via \eqref{Green} for an imaginary frequency
$\omega=iw$.  This is simply the inversion of a linear operator
$[\nabla\times\nabla\times+w^{2}\varepsilon(\vec{r},iw)]$ that has the
convenient properties of being real-symmetric and positive-definite,
as stated above.  Almost any technique developed for computational
electromagnetism is applicable here, modified to operate at an
imaginary frequency.  To illustrate our approach, we used a very
simple, yet extremely general, method: finite-difference
frequency-domain (FDFD) discretization of \eqref{Green} in a staggered
Yee grid~\cite{Christ87}, which we then invert by a conjugate-gradient
method~\cite{bai00}. Although the Yee grid in principle allows
second-order--accurate finite-difference approximations, unfortunately
the whole scheme becomes only first-order--accurate once a
discontinuous dielectric function $\varepsilon$ is included.  (There
are ways to treat interfaces more accurately~\cite{Farjadpour06}, but
we did not implement them here.)  Moreover, a very high resolution is
often required to resolve the stress tensor close to a dielectric
boundary due to the Green's function divergences as a boundary is
approached~\cite{Hackbush89}.  Despite its shortcomings, however, we
found FDFD to be sufficient to obtain accurate results (to within a
few percent in a reasonable time) for two-dimensional, and
three-dimensional $z$-invariant, geometries.  The key advantage of FDFD
is its flexibility: with very little effort, we were able to implement
support for arbitrary geometric shapes and arbitrary materials (both
perfect metals and dispersive/dissipative dielectrics).

Again choosing the simplest possible approach, we apply periodic
boundary conditions at the edges of the computational cell, which are
accurate as long as the boundaries are sufficiently far compared to
the separation between the interacting bodies.  That is, the
periodicity leads to artificial ``wrap-around'' forces that decay
rapidly with cell size $L$ (at least as $1/L^3$ in 2d); we chose cell
sizes large enough to make these contributions negligible ($< 1\%$).

\begin{figure}[hbt]
\includegraphics[width=0.35\textwidth]{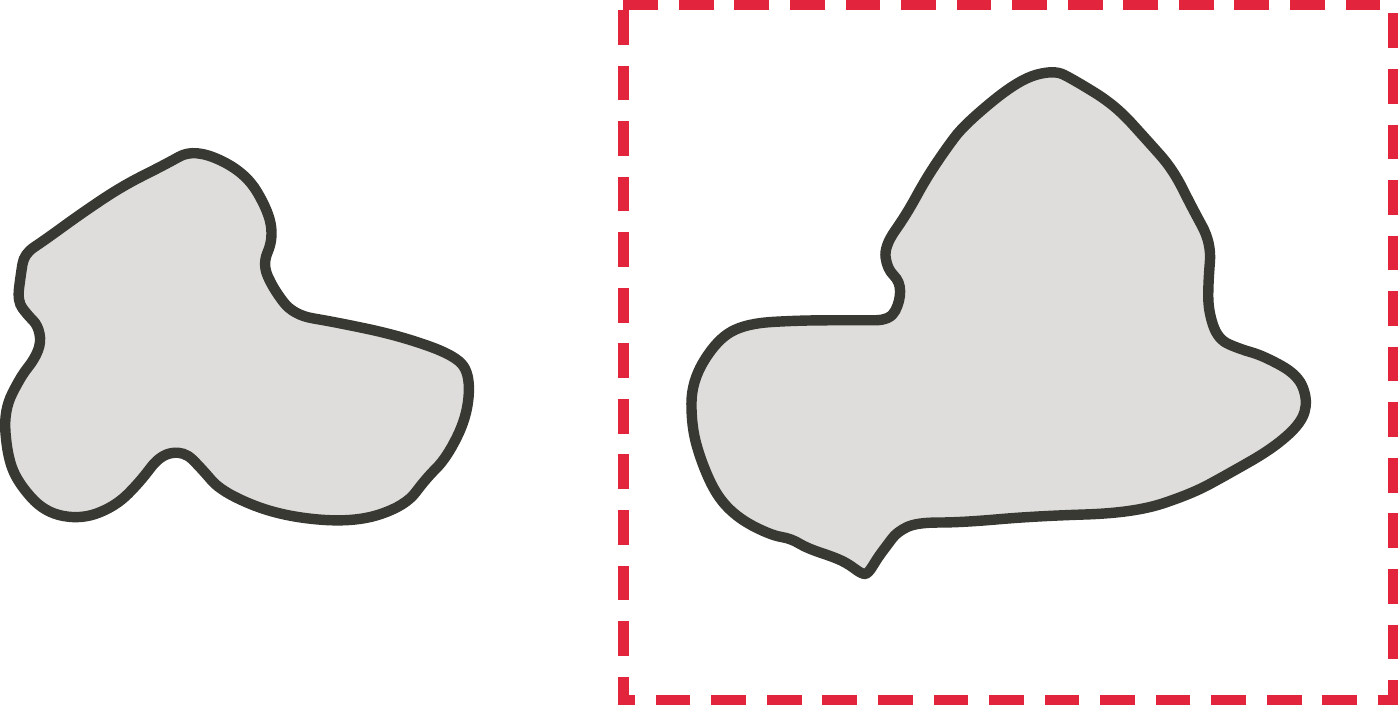}
\caption{(Color) Schematic illustration of a possible contour around a body;
the force on the body is given by an integral of the stress tensor
around this contour.}
\label{fig:contour}
\end{figure}

The computational process (using a simple finite-difference scheme)
goes as follows:

\begin{enumerate}
  \item Pick a contour/surface around the body of interest, as in
  \figref{contour} (which will typically \emph{not} coincide with the
  boundary of the body).
  \item For a given frequency $w$:
    \begin{enumerate}
      \item For every grid point $\vec{x}$ on the discretized contour/surface,
      solve \eqref{Green} for each polarization $k$ to obtain
      $G_{jk}(iw; \vec{x}-\vec{x})$.
      \item Integrate the resulting stress tensor $T_{jk}$
      over the surface, as in \eqref{F}.
    \end{enumerate}
  \item Integrate the above function over $w$ to obtain the force;
  since the integrand is a smooth function of $w$, an efficient
  adaptive quadrature scheme can be employed~\cite{piessens83}.
\end{enumerate}

Although this scheme does not require any additional regularization
(the integrand is finite for a finite resolution, and the integral of
the vacuum stress tensor over the contour is zero), we have found that
numerical convergence can be accelerated by subtracting the
stress-tensor integral over the isolated bodies.  For example, in the
schematic of \figref{contour}, we would first compute the
stress-tensor integral for the two bodies as shown, then subtract the
integral of the stress tensor over the \emph{same} surface with one
body removed, and then subtract again for the stress tensor with the
other body removed.  Of course, these subtracted quantities are zero
in the limit of infinite spatial resolution---there is no net force on
an isolated body.  However, at a finite resolution the discretization
error at the interface between two materials can lead to a finite
force that vanishes as resolution is increased.  By subtracting this
error term from the force, we find that the numerical error is
typically reduced by an order of magnitude or so. We emphasize,
however, that this is merely an optimization---even without
subtraction, the force converges to the correct result, and merely
requires a somewhat higher resolution.

\section{Forces in translation-invariant structures}
\label{sec:periodic-forces}

It is common to solve for the Casimir force between bodies that are
translation-invariant in one or more directions; for example, between
a cylinder and a plate~\cite{emig06} that are invariant in the $z$
direction.  More generally, one might consider structures that are
periodic in some direction with a non-zero period $\Lambda$, where
$\Lambda\rightarrow0$ corresponds to translation invariance.
Intuitively, in such cases one need only perform computations in the
unit cell of the periodicity, and the spatial integration in the
invariant direction(s) is replaced by integration over a wavevector
$\vec{k}$ from Bloch's theorem~\cite{Joannopoulos95, emig04_2}.
Although special cases of this familiar idea are well known in Casimir
computations~\cite{emig04_2, emig05}, here we provide a review of this
approach for an arbitrary periodicity in the context of stress-tensor
computational methods; the detailed derivation is provided in the
Appendix.  We also mention a useful optimization for the common
special case of perfect-metallic $z$-invariant structures.

Let us consider a single direction of periodicity: suppose that the
structure is periodic in $z$ with period $\Lambda$.  In this case, it
is natural to choose a surface for our stress-tensor integral that is
also periodic in $z$.  For example, imagine that \figref{contour}
depicts a two-dimensional ($xy$) cross-section of a $z$-invariant
structure, and the dashed line depicts a cross-section of the
corresponding $z$-invariant stress-tensor surface.  Because the total
force is infinite, the quantity of interest is the force per unit $z$.
It is convenient to consider the net force $\vec{F}$ from a finite
length $L = N\Lambda$ with periodic boundaries, and obtain the force
per unit length as $\lim_{N\rightarrow\infty} \vec{F}/L$.  Naively,
$\vec{F}/L$ can be written directly via \eqref{F}, where we break the
integral over $z$ into a summation over the unit cells:
\begin{align}
 \frac{\vec{F}}{L} &= \frac{1}{L} \int^{\infty}_0 dw \sum^{N-1}_{n=0} \iint 
 \mathbf{T}(iw; \vec{r}-n\Lambda\hat{\vec{z}}) \cdot \dS ,
\label{eq:Flength}
\\
 &= \frac{1}{\Lambda} \int^{\infty}_0 dw \iint 
 \mathbf{T}(iw; \vec{r}) \cdot \dS ,
\label{eq:Flengthtwo}
\end{align}
where the surface integral is over the portion of the surface lying in
the unit cell only, and in the second line we have used the fact that
the stress tensor is periodic.  This expression is inconvenient,
however, because the direct evaluation of $\mathbf{T}(iw, \vec{r})$
requires the response to a \emph{single} point source in the large-$L$
structure, and a single point source does \emph{not} produce a
periodic field (requiring a full three-dimensional calculation even
for a $z$-invariant structure).  Rather, one would like to consider
the field in response to \emph{periodic} point sources, which produce
a periodic field that can be treated by a small computational cell
with periodic boundary conditions.  This is accomplished by
Fourier-transforming the expressions in \eqref{Flength} and taking the
$N\rightarrow\infty$ limit, as described in detail by the Appendix.
The resulting force per unit length is:
\begin{equation}
 \frac{\vec{F}}{L} 
 = \frac{1}{\Lambda} \int^{\infty}_0 dw \int_{-\pi/\Lambda}^{\pi/\Lambda} \frac{dk_z}{2\pi} \iint 
 \mathbf{T}(iw, k_z; \vec{r}) \cdot \dS ,
\label{eq:Freduced}
\end{equation}
where the surface integral is still over the portion of the surface
lying in the unit cell. Here, $\mathbf{T}(iw, k_z; \vec{r})$ denotes
the stress tensor computed from the Green's functions for
Bloch-periodic boundaries---that is, from the fields in response to a
periodic set of point-dipole sources with phase $e^{ik_z\Lambda n}$ in
the $n$th unit cell.  This stress tensor can be computed using a
computational cell that is only one unit cell in the $z$ direction,
e.g. by a two-dimensional computational cell for a $z$-invariant
structure.  [Equivalently, $\mathbf{T}(iw, k_z; \vec{r})$ could
instead be computed from the Green's function for ordinary periodic
boundaries, but with $\nabla$ replaced by $\nabla+ik_z
\hat{\vec{z}}$~\cite{Joannopoulos95}.]  Just as for $w$, the stress
tensor is a smooth function of $k_z$ and therefore the $k_z$ integral
can be computed by an efficient quadrature scheme (e.g. Gaussian
quadrature).

If the structure is periodic (or invariant) in more than one
direction, one simply repeats the above procedure: for each periodic
direction, we only consider the portion of the stress-tensor integral
in the unit cell, with Bloch-periodic boundary conditions, and
integrate over the corresponding Bloch wavevector component.  Also, by
symmetry one only need integrate over the irreducible Brillouin zone
of the structure~\cite{Joannopoulos95}, e.g. in one dimension (where
time-reversal symmetry normally equates $k_z$ and $-k_z$) the
$\int_{-\pi/\Lambda}^{\pi/\Lambda} dk_z$ integral can be replaced by
$2 \int_0^{\pi/\Lambda} dk_z$.

For the common case of a $z$-invariant perfect-metal structure
(i.e. one has a homogeneous $\varepsilon$ surrounded by perfect
metal), there are several important simplifications.  First, the
$(k_z, w)$ Green's function is \emph{exactly} the same as the $(0,
\sqrt{k_z^2 + w^2})$ solution~\cite{emig06}.  Therefore, we need only
compute the $k_z=0$ solutions at each $w$, and weight the $dw$
integrand by a factor of $\pi w$ (the circumference of a semi-circle
of radius $w$) divided by the $2\pi$ that would appear in the $dk_z$
integral. (For two directions of translational symmetry, one would
weight the integral by the area of a hemisphere of radius $w$, and in
general, in $d$ dimensions, by the area of a radius-$w$ hypersphere.)
Furthermore, at $k_z=0$, the solutions can be divided into two scalar
polarizations, TE ($\vec{E}\cdot\hat{\vec{z}}=0$) and TM
($\vec{H}\cdot\hat{\vec{z}}=0$)~\cite{Jackson98}.

\section{Numerical Results and Visualization}
\label{numerical}

In the following sections we demonstrate our method's validity by
checking it against known results for perfect metals, and in
particular for the case of a cylinder adjacent to a plate. A new
geometry that displays an interesting non-monotonic behavior is
presented for both perfect and realistic dispersive metals, and in
both 2d and 3d. Furthermore, we explain how one can use stress-tensor
maps to visualize the interactions between bodies and identify the
most important spatial regions.

\subsection{Parallel plates in one dimension}

\begin{figure}[hbt]
\centerline{\includegraphics[width=0.5\textwidth]{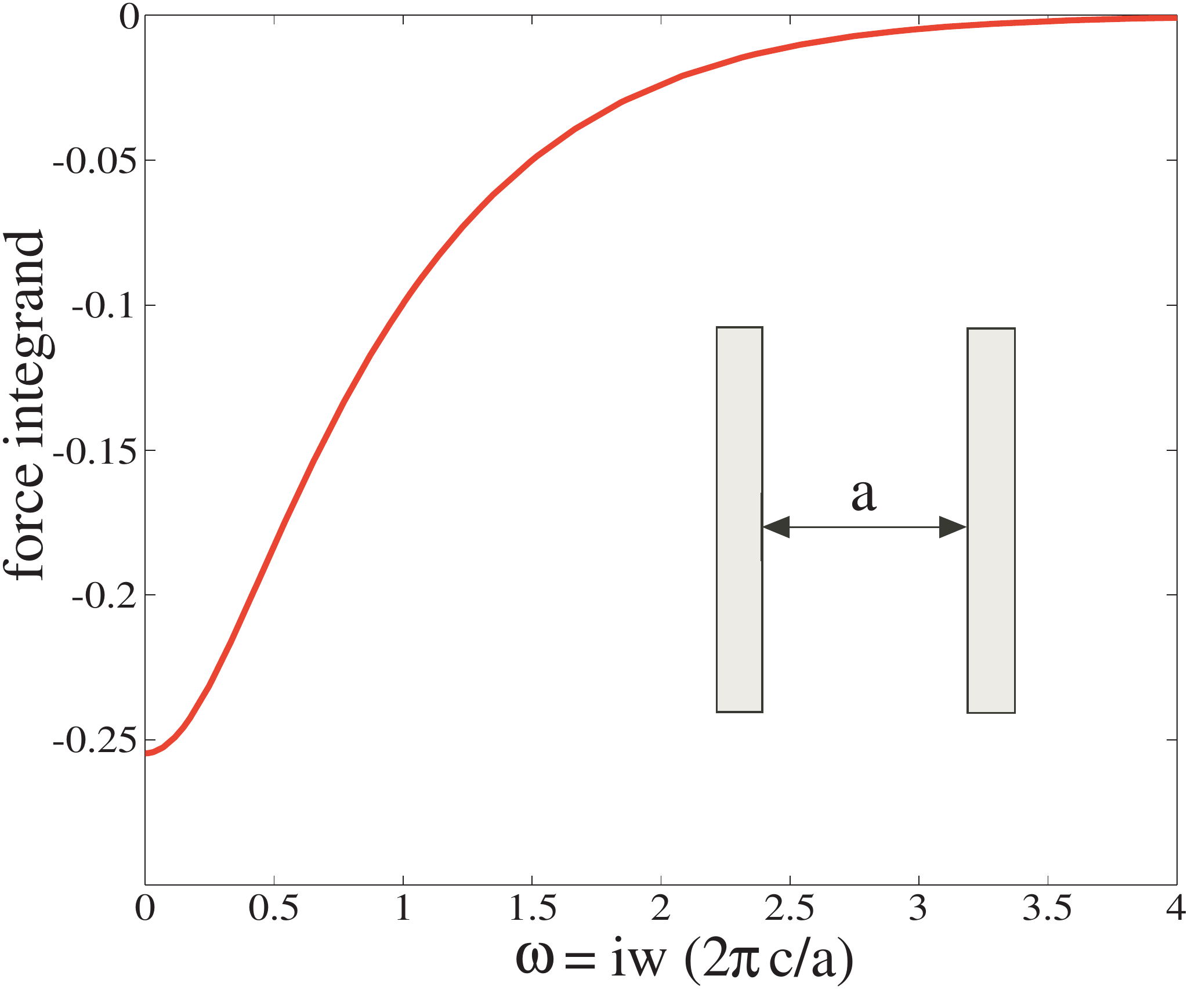}}
\caption{(Color) Plot of Casimir force integrand $\oiint T \cdot
\dS$ between two 1d parallel plates separated by a distance $a=1$
versus imaginary frequency $w = \Im \omega$, using the stress-tensor
method of \secref{stress-approach}.}
\label{fig:stress-integrand}
\end{figure}

First, for comparison to the one-dimensional parallel-plate integrands
plotted in \figref{fdensity} and \figref{psum-casimit}, we plot the
corresponding integrand for the stress tensor in
\figref{stress-integrand}.  Again, this integrates to the correct
force ($\pi \hbar c / 24 a^2$ for each polarization), but
we note that the integrand is not identical to the integrand from the
imaginary-frequency energy derivative.

Since the one-dimensional parallel-plate force is commonly derived
from the Lifshitz formula, which in turn is derived from the stress
tensor, this cannot be regarded as a rigorous validation of our method
(except in the most basic sense of checks for bugs in our code).

\subsection{Cylinder and plate}

\begin{figure}[hbt]
\centerline{\includegraphics[width=0.5\textwidth]{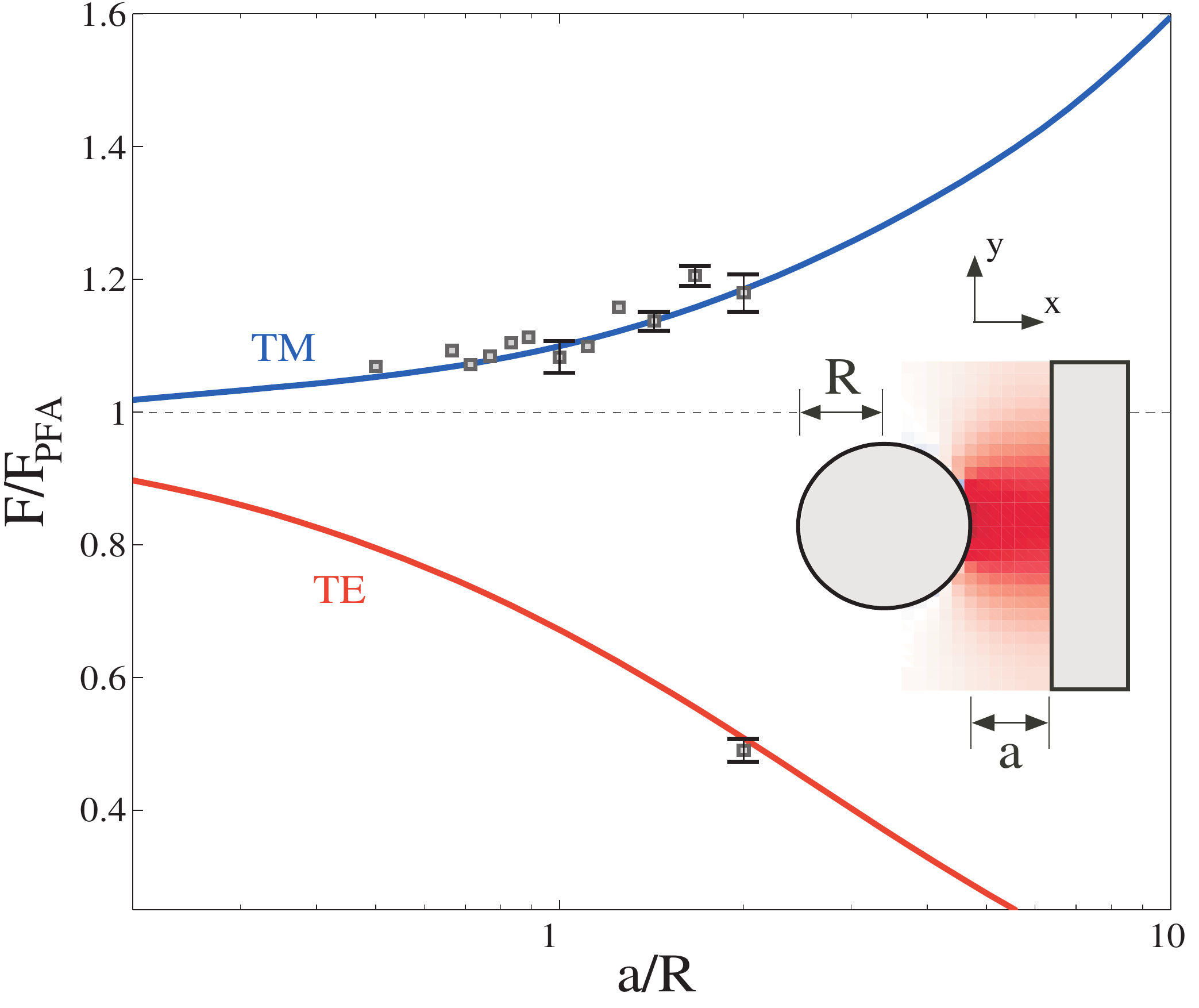}}
\caption{(Color) Casimir force between a 3d radius-$R$ cylinder and a plate
(inset), relative to the proximity-force approximation
$F_\mathrm{PFA}$, vs. normalized separation $a/R$. The solid lines are
the Casimir force computed in \citeasnoun{emig06} for TE (gray) and TM
(blue) polarizations, along with results computed by our method with a
simple finite-difference discretization (gray squares). Error bars were
estimated for some data points by using computations at multiple
spatial resolutions.  Inset shows interaction stress tensor $\Delta
\langle T_{xx} \rangle$ at a typical imaginary frequency $w =
2\pi c/a$, where red indicates attractive stress.}
\label{fig:cylinder}
\end{figure}

A more complicated geometry, consisting of a perfect metallic cylinder
adjacent to a perfect metallic plate in three dimensions, was solved
numerically by \citeasnoun{emig06}, to which our results are compared
in \figref{cylinder}.  \citeasnoun{emig06} used a specialized
Fourier-Bessel basis specific to this cylindrical geometry, which
should have exponential (spectral) convergence. Our use of a simple
uniform grid was necessarily much less efficient, especially with the
first-order accuracy, but was able to match the \citeasnoun{emig06}
results within $ \sim 3\%$ using reasonable computational resources. A
simple grid has the advantage of being very general, as illustrated
below, but other general bases with much greater efficiency are
possible using finite-element or boundary-element methods; the latter,
in particular, could use a spectral Fourier basis similar to
\citeasnoun{emig06} and exploit a fast-multipole method or similar
$O(N \log N)$ solver technique.  Surface discretizations (boundary
elements) will also have the advantage that the infinite amount of space
surrounding the objects is treated analytically rather than having to
be truncated with some boundary conditions (here, periodic).  This is
discussed in greater detail in \secref{beyond-FDFD}.

Also shown, in the inset of \figref{cylinder}, is a plot of the
interaction stress-tensor component $\Delta \langle T_{xx}
\rangle$ at a typical imaginary frequency $w = 2\pi c/a$.  By
``interaction'' stress tensor $\Delta \langle T_{ij}\rangle$,
we mean the total $\langle T_{ij}\rangle$ of the full
geometry minus the sum of the $\langle T_{ij}\rangle$'s
computed for each body in isolation.  Here, the stress tensors of the
isolated cylinder and plate have been subtracted, giving us a way to
visualize the force due to the interaction.  As described further
below, such stress plots reveal the spatial regions in which two
bodies most strongly affect one another, and therefore reveal where a
change of the geometry would have the most impact.  (In contrast,
\citeasnoun{gies06:edge} plots an interaction-energy density that does not
directly reveal the force, since the force requires the energy to be
differentiated with respect to $a$.  For example,
\citeasnoun{gies06:edge}'s subtracted energy density apparently goes nearly to
zero as a metallic surface is approached, whereas the stress tensor
cannot since the stress integration surface is arbitrary.)


\subsection{Two-dimensional metal piston and non-monotonic ``lateral'' forces}

\begin{figure}[hbt]
\centerline{\includegraphics[width=0.5\textwidth]{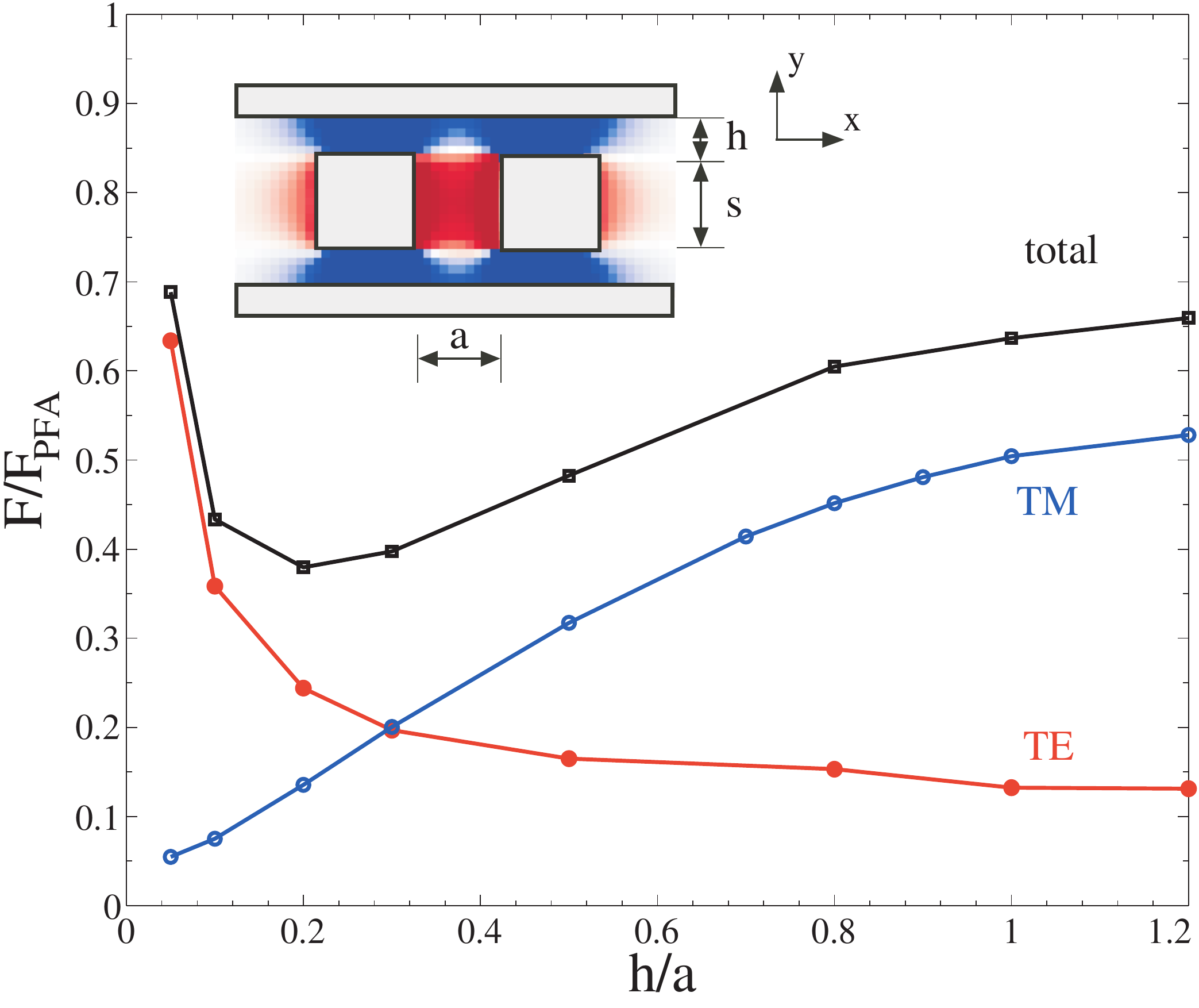}}
\caption{(Color) Casimir force between 2d ($z$-invariant fields) metal squares
$F/F_\textrm{PFA}$, vs. distance from metal plate $h$ (inset),
normalized by the total force (TE+TM) obtained using the PFA,
$F_\textrm{PFA}=~\hbar c \zeta(3)s/8\pi a^3$. The total force is
plotted (black squares) along with the TE (red dots) and TM (blue
circles) contributions.}
\label{fig:metal-blocks}
\end{figure}

We now consider a more complicated geometry in which there are
interactions between multiple bodies: a two-dimensional
``piston''-like structure, shown in \figref{metal-blocks}, consisting
of two metal $s \times s$ squares separated by a distance $a$ from one
another (here, $s = a$) and separated by a distance $h$ from infinite
metal plates on either side.  We then compute the Casimir force
between the two squares, in two dimensions (that is, for $z$-invariant
fields, unlike the cylinder case above where $z$ oscillations were
included), as a function of the separation $h$.  The result for
perfect conductors is shown in \figref{metal-blocks}, plotted for the
TE and TM polarizations and also showing the total force.  (Error bars
are not shown because the estimated error is $< 1\%$.  This structure
is computationally easier than the cylinder-plate problem, for a
finite-difference discretization, because the metallic walls allow the
computational cell size to be small in at least one direction.) In the
limit of $h\rightarrow 0$, this structure approaches the 2d ``Casimir
piston,'' which has been solved analytically for the TM
polarization~\cite{cavalcanti04}.  Our results, extrapolated to $h=0$,
agree agree with this analytical result to within $3\%$ (although we
have computational difficulties for small $h$ due to the high
resolution required to resolve a small feature in FDFD).  For $h > 0$,
however, the result is surprising in at least two ways.  First, the
total force is \emph{non-monotonic} in $h$, due to a competition
between the TE and TM contributions to the forces.  Second, the $h$
dependence of the force is a \emph{lateral} effect of the parallel
plates on the squares, which would be zero by symmetry in PFA or any
other two-body--interaction approximation.

The reader may notice that the TE and TM forces in the cylinder-plate
case, \figref{cylinder}, also have opposite-sign slopes in the graph,
and one may therefore suspect that non-monotonic forces could occur in that
case as well.  However, in the cylinder-plate case this apparent
difference in sign is merely an artifact of the normalization: the PFA
force varies with the separation $a$, and when the actual force (which
is monotonically decaying with $a$) is divided by this variable force
one can obtain ``non-monotonic'' plots.  (A similar ``non-monotonic''
plot can be seen in \citeasnoun{emig06}.)  In contrast, for
\figref{metal-blocks}, the PFA normalization is constant because $a$ is
fixed, and thus the relative forces from different $h$ are directly
comparable.

Although lateral forces can still arise qualitatively in various
approximations, such as in ray optics or in PFA restricted to
``line-of-sight'' interactions, it may not be immediately clear how
these could predict non-monotonicity.  We also note that, in the
large-$h$ limit, the force remains different from PFA due to
finite-$s$ ``edge'' effects~\cite{gies06:edge}, which are captured by
our method.  It turns out that one can qualitatively predict the
non-monotonic behavior, due to the competition between TE and TM
forces, using the ray-optics approximation, although this
approximation is not quantitatively accurate except for $h=0$; we will
describe this ray-optics analysis in a future
publication~\cite{Jaffe07:ray}.

\begin{figure}[hbt]
\includegraphics[width=0.5\textwidth]{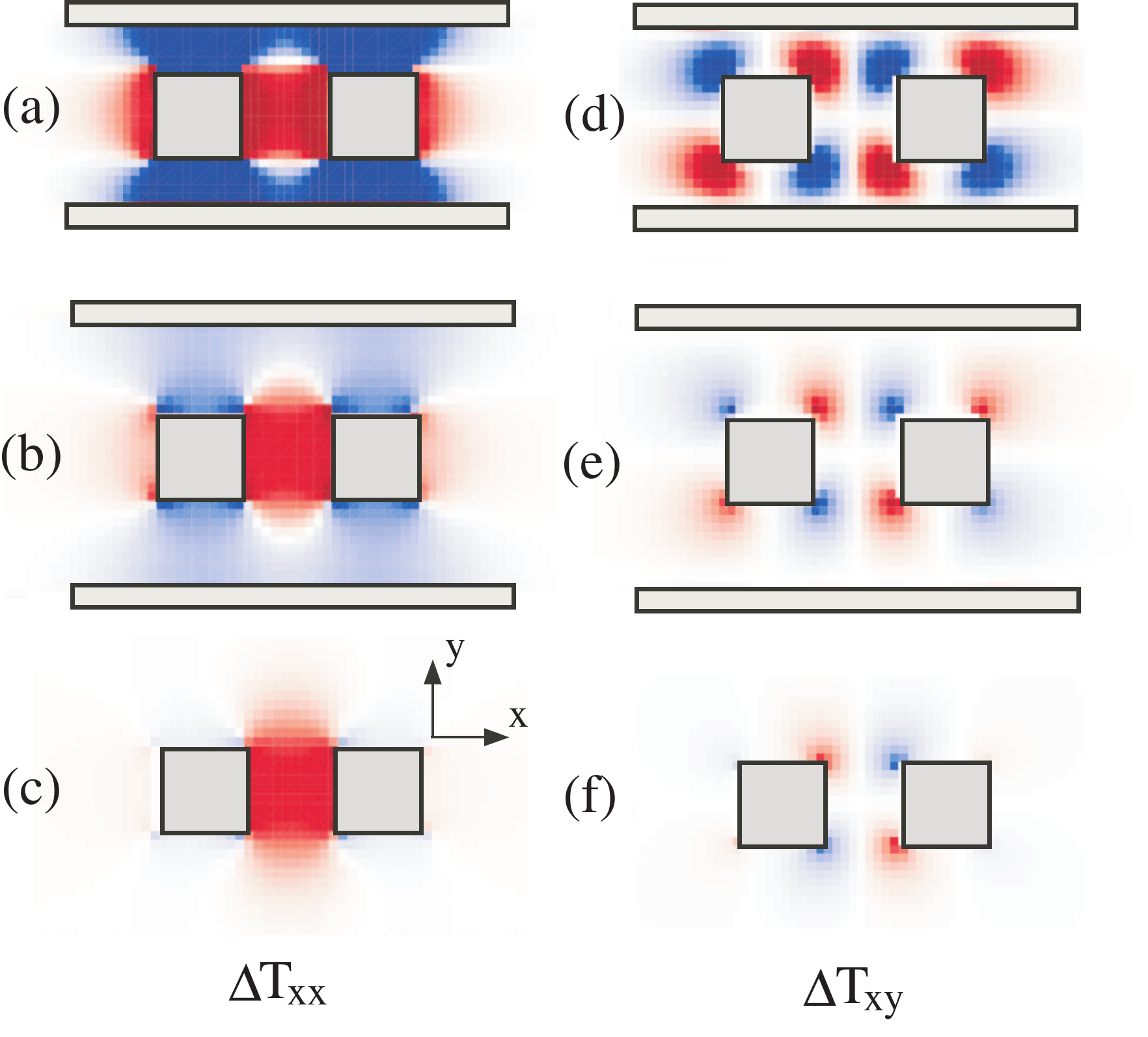}
\caption{(Color) (a--f): TM stress map of the geometry in
\figref{metal-blocks} for various $h$.  The interaction stress tensors
$\langle T_{xx} \rangle$ (left) and $\langle T_{xy}
\rangle$ (right) for: (a),(d): $h = 0.5a$; (b),(e): $h = a$; and
(c),(f): $h = 2a$, where blue/white/red = repulsive/zero/attractive.}
\label{fig:sub}
\end{figure}

To further explore the source of the $h$-dependence, we plot the TM
interaction-stress maps $\Delta \langle T_{xx} \rangle$ and
$\Delta \langle T_{xy} \rangle$ in \figref{sub}, for the
perfect-metal squares at a typical frequency $w = 2\pi c / a$, and for
varying distances from the metal plates ($h = 0.5$, $1.0$, $2.0$). As
shown, the magnitudes of both the $xx$ (a--c) and $xy$ (d--f)
components of the stress tensor change dramatically as the metal
plates are brought closer to the squares. For example, one change in
the force integral comes from $T_{xy}$, which for isolated
squares has an asymmetric pattern at the four corners that will
contribute to the attractive force, whereas the presence of the plates
induces a more symmetric pattern of stresses at the four corners that
will have nearly zero integral.  This results in a decreasing TM force
with decreasing $h$ as observed in \figref{metal-blocks}.  Because
stress maps indicate where bodies interact and with what signs, it may
be useful in future work to explore whether they can be used to design
unusual behaviors such as non-additive, non-monotonic, or even
repulsive forces.

\subsection{Two-dimensional dielectric pistons}

Our method is also capable, without modification, of handling
arbitrary dielectric materials.The calculation of general dispersive
media can be performed with minor or no additional computational
effort, since the computations at different $w$ are
independent. Unlike most previously published techniques, which do not
easily generalize to non-perfect metals, the stress tensor approach
does not distinguish between the two regimes, and computational
methods for inhomogeneous dielectric materials are widely available.
Furthermore, we reiterate that along the imaginary-$\omega$ axis,
$\varepsilon$ is purely real and positive even for dissipative
materials (which have complex $\varepsilon$ on the real-$\omega$
axis), greatly simplifying computations.

The method's ability to handle dielectric structures is demonstrated
below, where the Casimir force between the two
squares is shown for two different cases: 

\begin{figure}[ht]
\centerline{\includegraphics[width=0.5\textwidth]{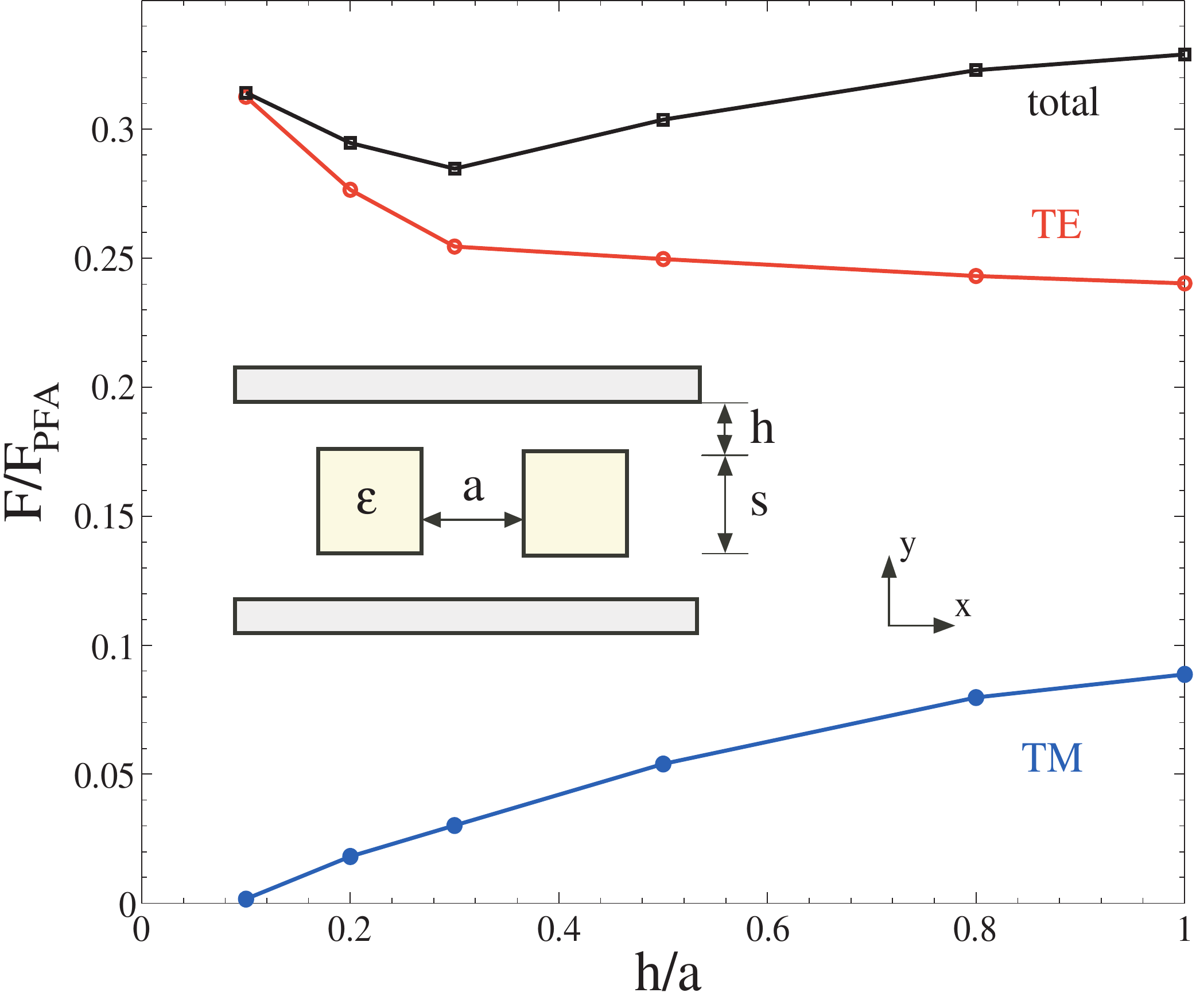}}
\caption{(Color) Casimir force between 2d ($z$-invariant fields) dielectric
($\varepsilon=4$) squares $F/F_\textrm{PFA}$, vs. distance from metal
plate $h$ (inset), normalized by the total force (TE+TM) obtained
using the PFA (Here, the PFA force is computed for $x$-infinite slabs
of dielectric $\varepsilon=4$). The total force is plotted (black
squares) along with the TE (red dots) and TM (blue circles)
contributions.}
\label{fig:eps-blocks}
\end{figure}

\begin{figure}[ht]
\centerline{\includegraphics[width=0.5\textwidth]{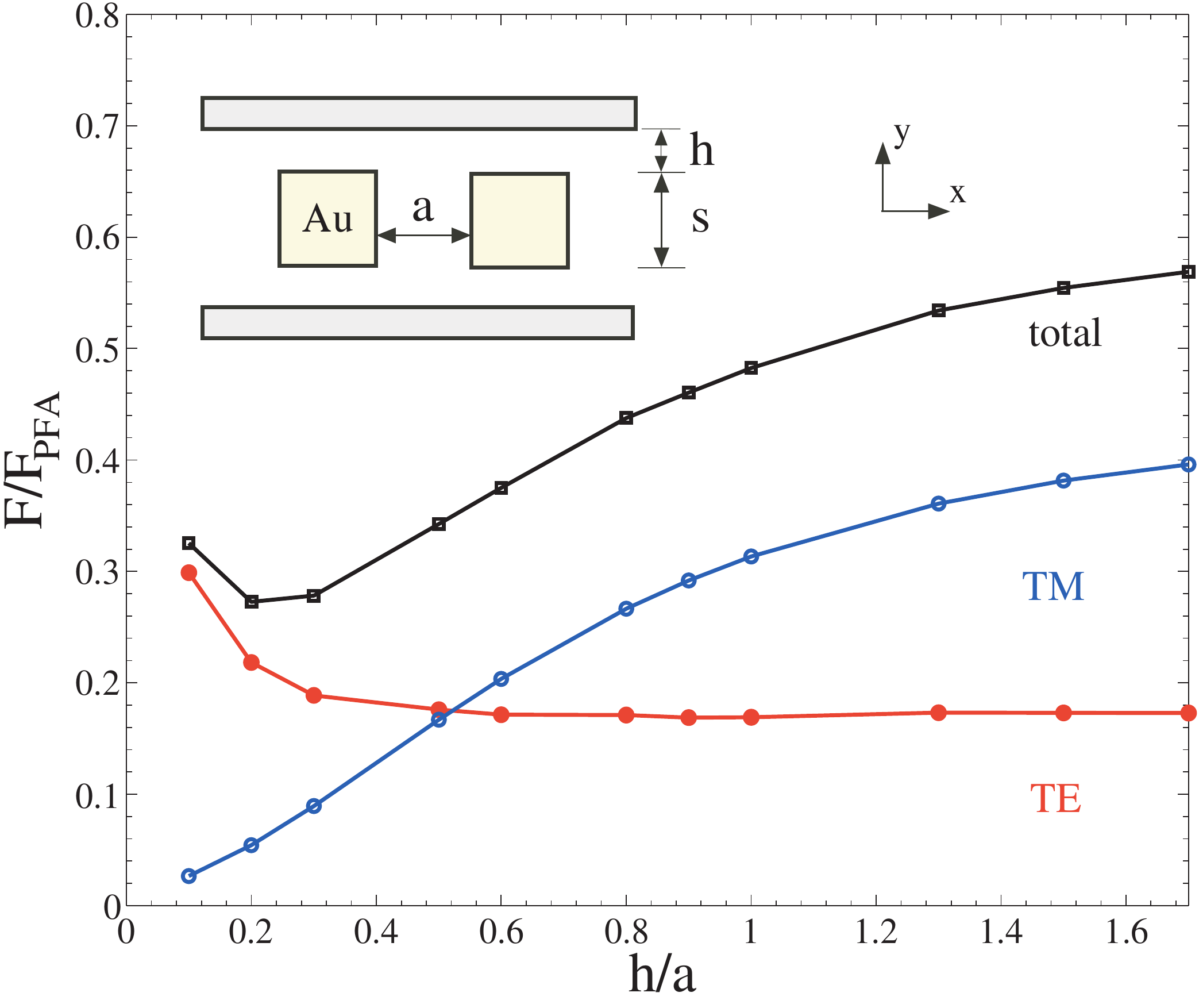}}
\caption{(Color) Casimir force between 2d ($z$-invariant fields) gold squares
$F/F_\textrm{PFA}$, vs. distance from metal plate $h$ (inset),
normalized by the total force (TE+TM) obtained using the PFA. (Here,
the PFA force is computed for $x$-infinite gold slabs). The total
force is plotted (black squares) along with the TE (red dots) and TM
(blue circles) contributions.}
\label{fig:gold-blocks}
\end{figure}

First, we compute the force between two squares made of dielectric
material with $\varepsilon = 4$ (an artificial mathematical choice for
illustration purposes), whereas the parallel plates are still perfect
metal.  The result is shown on the plot of \figref{eps-blocks}. As
might be expected, the dielectric squares have a weaker interaction
than the perfect-metal squares, but are still non-monotonic.

Second, as a more interesting example, the squares are made of gold
with a Drude dispersion taken from experiment, again with adjacent
perfect metallic plates. In particular, the following Drude model is
used for the material dispersion of gold~\cite{Brevik05}:
\begin{equation}
\varepsilon(\omega) =
1-\frac{\omega^2_p}{\omega\left(\omega+i\Gamma_p\right)}
\end{equation}
with $\omega_p=1.37\times10^{16}$~Hz and
$\Gamma_p=5.32\times10^{13}$~Hz, corresponding to $\omega_p=7.2731$
and $\Gamma_p = 0.028243$ in our units of $2\pi c/a$, for $a=1$
$\mu$m.  For $\omega=iw$, this is real and positive, as expected.  The
resulting force is shown in \figref{gold-blocks}. Not surprisingly,
the gold squares have a weaker interaction than perfect-metal squares,
since at large $w = \Im \omega$ the dielectric constant $\varepsilon$
goes to $1$.

\subsection{Three-dimensional piston}

\begin{figure}[hbt]
\centerline{\includegraphics[width=0.5\textwidth]{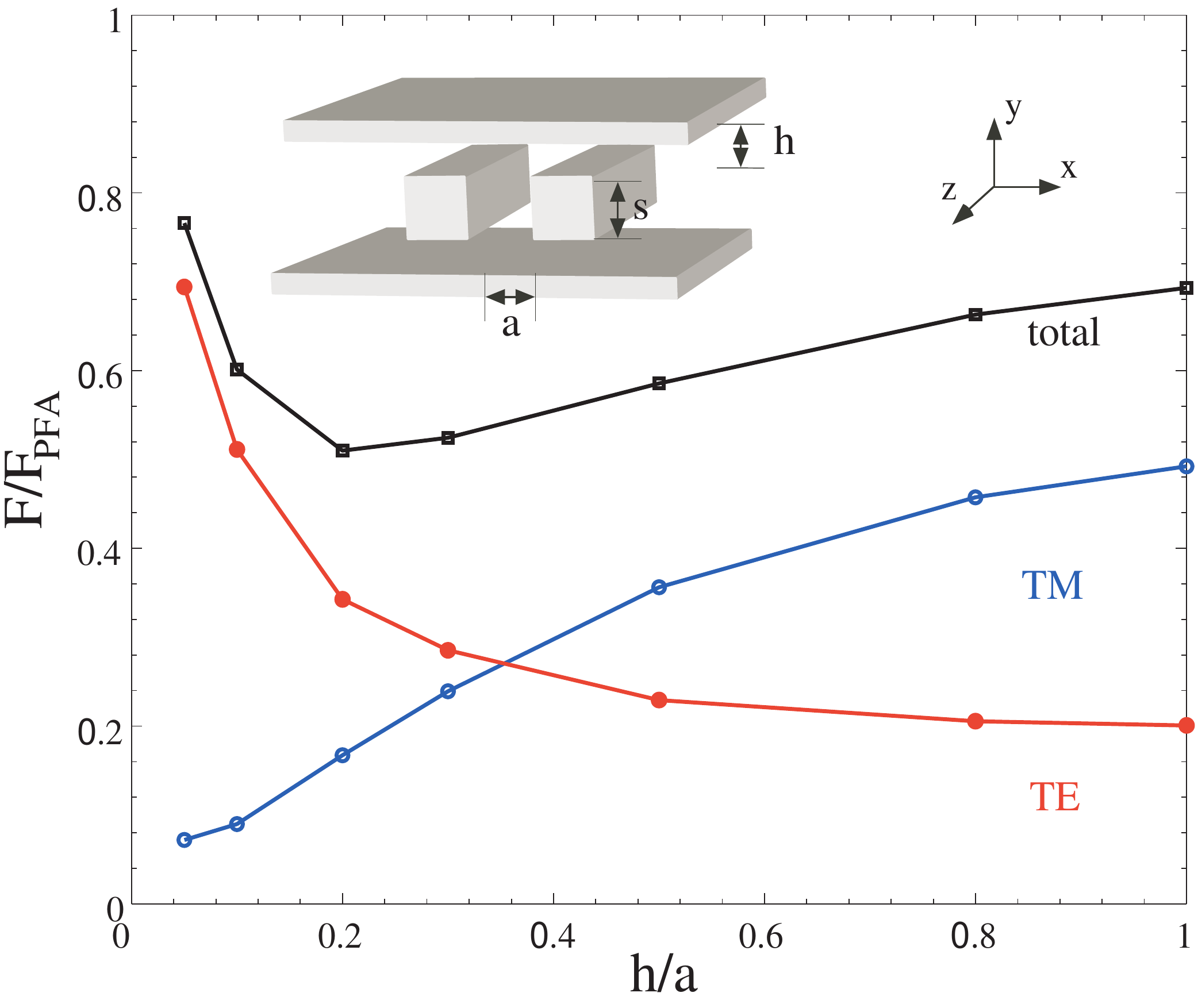}}
\caption{(Color) Casimir force per unit length between $z$-invariant 3d metal blocks
$F/F_\textrm{PFA}$, vs. distance from metal plate $h$ (inset),
normalized by the total force (TE+TM) obtained using the PFA,
$F_{\textrm{PFA}}~=~\hbar c s \pi^2 / 480 a^4$. The total force is
plotted (black squares) along with the TE (red dots) and TM (blue
circles) contributions.}
\label{fig:3d-force}
\end{figure}

The previous 2d calculations are important in at least two ways:
first, they allow us to check the stress tensor method against
previous piston calculations in the $h\rightarrow 0$ limit, while
exploring an interesting new geometry; second, based on their results,
one might predict a similar behavior for the force per unit length in
the analogous 3d $z$-invariant geometry, shown in the inset of
\figref{3d-force}. Indeed, this is the case: the non-monotonic force
in three dimensions is shown in \figref{3d-force}. As discussed in
\secref{periodic-forces}, the integrand for a $z$-invariant
perfect-metallic structure differs from the two-dimensional integrand
only by a factor of $\pi w$, and therefore the 3d force is obtained
from the 2d calculations with very little additional computation.

Again, in the $h=0$ limit there are known analytical solutions for this
geometry based on the ray-optics method~\cite{Hertzberg05} or the
zeta-function technique~\cite{Marachevsky07}.  Linearly extrapolating
our plot to $h=0$, we reproduce these results to within $\approx
2$--$3\%$.

\section{Beyond finite-difference methods}
\label{sec:beyond-FDFD}

Above, we implemented the stress-tensor integration using a
finite-difference frequency-domain approach to compute the Green's
function.  While sufficient for a proof-of-concept implementation, one
would like to use more sophisticated methods in order to explore
complex geometries more quickly, especially in three dimensions.  The
primary drawbacks of the finite-difference scheme are threefold.
First, the material discontinuities imply that the error converges
only linearly with resolution, although there are techniques to
improve this to quadratic convergence~\cite{Farjadpour06}.  Second,
non-uniform resolution would be desirable to handle small features,
such as the narrow channels in the piston structure for small $h$.
Third, the stress-tensor integrand is a discretized, non-smooth
function of space, meaning that we must evaluate it at a number of
grid points proportional to the resolution (or resolution squared, in
three dimensions); in general, the integrand must therefore be
evaluated $O(N^{(d-1)/d})$ times for a $d-1$ dimensional surface in
$d$ dimensions, leading to at best $O(N^{2-1/d})$ complexity.  To
address these drawbacks, we consider two standard approaches to
solving partial differential equations in a more efficient manner in
complex geometries: finite-element and boundary-element methods.

Finite-element methods can employ a non-uniform volume discretization,
via an unstructured mesh, that can both put more resolution where it
is needed and conform to the material interfaces to obtain
higher-order accuracy~\cite{chew01,Volakis01,Jin02,Zhu06}.  However,
from the perspective of Casimir-force calculations, finite-element
methods seem to have two potential drawbacks.  First, because space is
still discretized, the stress tensor is again not a smooth function of
space and its accurate integration requires that the Green's function
be evaluated at many mesh points.  Second, there may be a problem with
regularization: although the Green's function does not diverge in
discretized space, with a non-uniform resolution this effective
regularization varies at different points.  Unless there is a way to
locally regularize the problem (subtracting a vacuum Green's function
computed at the local spatial resolution), this may lead to
unphysical, non-convergent forces.

Boundary-element methods (BEMs) involve a discretization in terms of
unknowns only at the interfaces between different materials---these
surface unknowns are coupled to one another via the (known) Green's
functions of the homogeneous
regions~\cite{Hackbush89,chew97,Rao99,chew01,Volakis01,Jin02}.  They
are thus ideal for open problems, in which the bodies are surrounded
by infinite volumes of empty space, since those infinite regions are
treated analytically.  Using the fast-multipole method (FMM) or other
fast integral-equation methods~\cite{Hackbush89, chew01, Volakis01,
Jin02}, BEMs can solve for the surface unknowns, and hence the Green's
function, in $O(N \log N)$ time for $N$ discretized unknowns,
multiplied by a number of iterations ($\ll N$) that depends on the
condition number of the matrix and the
preconditioning~\cite{bai00}. Furthermore, BEMs have two unique
advantages when applied to the problem of Casimir forces.  First, the
regularization can be performed analytically: since BEMs express the
Green's function as the sum of the vacuum Green's function plus a set
of contributions from surface currents, the vacuum Green's function
can be trivially subtracted analytically.  Second, because space is
not discretized, the stress tensor in a BEM will be a \emph{smooth}
(infinitely differentiable) function of space---this means that the
spatial integral can be performed with exponentially-convergent smooth
quadrature methods.  Therefore, the number of times that the Green's
function must be computed is determined only by the convergence of the
smooth multidimensional quadrature in $2+1$ dimensions (space +
frequency), \emph{independent} of $N$.

For these reasons, we suspect that BEMs will lead to the most
efficient methods to compute Casimir forces for complicated structures
in three dimensions.  Moreover, all that needs to be done is to take
an existing BEM Green's function solver and change it to solve for the
imaginary-$\omega$ Green's function.  Because the imaginary-$\omega$
Green's functions are exponentially decaying (and approach the
familiar Poisson kernel as $\omega\rightarrow 0$), such a fast solver
should actually be simpler than the corresponding real-$\omega$
solver, nor are the singularities in the Green's function any worse.
And, as mentioned previously, the resulting matrix equation is
real-symmetric and positive-definite for imaginary $\omega$, unlike
the real-$\omega$ case, even for dissipative materials.  In short,
there do not appear to be any substantial unsolved algorithmic
problems involved in implementing a BEM for Casimir forces.

\section{Comparison to other methods}
\label{sec:comparison}

In this section, we briefly compare the stress-tensor approach with
other known exact numerical methods applicable to arbitrary
geometries, focusing mostly on the computational aspects. In
particular, we examine the methods of Emig~\textit{et
al.}~\cite{emig04_1} and Gies~\textit{et al.}~\cite{gies03}.

Emig's method, applicable to both separable and non-separable
geometries and not limited to perfect metallic structures (although
currently only demonstrated in perfect metallic separable geometries),
involves a surface parameterization of the Green's
function. Specifically, the Casimir energy is given in terms of an
integral over imaginary frequencies of the change in the photon
density of states (DOS) $\delta \rho(iw)$:
\begin{equation}
  U = \int \frac{\hbar w}{2}\delta \rho(iw) dw,
\end{equation}
similar to the expression in \eqref{energy-DOS}. However, the crucial
aspect of this method lies in the evaluation of the DOS, given
by~\cite{emig04_1}:
\begin{equation}
  \delta \rho (iw) = \frac{1}{\pi} \frac{\partial }{\partial w} \tr \ln \left(M^{-1}_{\infty} M\right).
  \label{eq:delta-rho}
\end{equation}
Here $M$ is an $N\times N$ dense matrix, where $N$ is the number of
surface degrees of freedom, whose entries are in terms of the
imaginary-$\omega$ vacuum Green's function $\vec{G}(iw;
\vec{x},\vec{x'})$ evaluated on the surface of each body. $M_\infty$
is the same matrix for the case where the bodies are infinitely far
apart.  Thus, the trace in \eqref{delta-rho} is analogous to an
integration over the surfaces of all the bodies.  Inverting a dense
matrix, multiplying two dense matrices, and taking the log of a dense
matrix all require $O(N^3)$ time (for practical algorithms) and
$O(N^2)$ storage~\cite{Golub96}.

We should comment however, that Emig's method is closely related to a
boundary-element method (BEM) as discussed in the previous section.
BEMs also involve parameterization in terms of surface degrees of
freedom, which are also coupled in terms of vacuum Green's functions,
leading to a dense matrix which must be inverted to compute the
inhomogeneous Green's function.  By recognizing this relationship, one
should be able to exploit fast-multipole and similar $O(N \log N)$
techniques to accelerate Emig's method.  In particular, computations
such as $M_\infty ^{-1} M$ involve the solution of $N$ linear
equations, each of which can employ iterative methods with $O(N)$
storage, and similar iterative methods may also be available for
computing matrix logarithms~\cite{Golub96}.  At best, this leads
to $O(N^2 \log N)$ time.  However, this is still much less efficient
than the stress-tensor BEM approach discussed in the previous section,
because the latter requires only $\ll N$ linear equations to be solved
(the number of linear equations to be solved is determined by a smooth
spatial quadrature, independent of $N$).

A second exact computational method available is that of
\citeasnoun{gies03}, based on a ``worldline'' approach. In this
method, the Casimir energy is represented via a scalar field in a
smooth background potential, and the effective action is obtained via
a Feynman path integral over all proper time worldlines using a
Monte-Carlo approximation.  The method has only been formulated for
the TM polarization with perfect-metal bodies, although preliminary
generalizations have been suggested~\cite{Gies06:worldline}.
Specifically, the method expresses the Casimir energy between two
bodies as the integral of a functional $\Theta_\Sigma$ over closed
paths $\vec{x}(\tau)$ of length (proper time) $T$ and center of mass
$\vec{y}$:
\begin{equation}
  U = -\frac{1}{8\pi^2} \int^{\infty}_0 \frac{dT}{T^3} \int d^3\vec{y} \langle \Theta_{\Sigma}[\vec{x}(\vec{\tau})]\rangle_{\vec{x}}
\end{equation}
where $\Theta_{\Sigma}[\vec{x}(\vec{\tau})]$ is a worldline functional
(similar to a step function) defined to be
$\Theta_{\Sigma}[\vec{x}(\vec{\tau})]=1$ if the path
$\vec{x}(\vec{\tau})$ intersects a surface and $0$
otherwise~\cite{Gies06:worldline}.  This integration is performed by
generating an ensemble of $n_L$ random $N$-point paths $\vec{x}(\tau)$
and evaluating $\Theta_\Sigma$ for each one.  The computational
complexity $O(n_L \cdot N \cdot \#\vec{y} \cdot \#T)$ (where
$\#\vec{y}$ and $\#T$ are the number of $\vec{y}$ and $T$ integrand
evaluations, respectively), therefore, depends on quantities such as
the precise statistical rate of convergence of this integral (error
$\sim 1/\sqrt{n_L}$), which in turn depends on the geometry and on the
manner in which the path ensembles are generated and integrated.
\citeasnoun{gies03} does not present a general analysis of these
quantities, nor will we do so here.  However, in the specific case of
the force between a radius-$R$ sphere and a plate separated by a
distance $a$, \citeasnoun{gies03} shows that $N \gg a^2 / R^2$ for
large $a/R$.  This does not include the $O(n_L \cdot \#\vec{y} \cdot
\#T)$ factors, although it seems likely that $\#\vec{y}$ is at least
$\sim N$ (so that the spatial resolutions are comparable), in which
case the time scaling would be at least $a^4/R^4 \sim N^2$.  In
comparison, a BEM for the same geometry (exploiting the cylindrical
symmetry to reduce it to a 2d problem similar to \citeasnoun{gies03})
should require degrees of freedom $N$ that never scale worse than
linearly with the relevant lengthscale, and time that scales with $N
\log N$ rather than $N^2$.  A general comparison seems difficult,
however, and a detailed statistical study is outside the scope of this
paper.

\section{Concluding remarks}
\label{sec:conc}

The general considerations involved in designing a purely
computational method are often quite different from those involved in
designing an analytical method.  For this reason, we believe it is most
fruitful to start back at the earliest possible formulations and
proceed using the new computational perspective, rather than
attempting to add more and more corrections to analytical methods for
specific geometries.  Moreover, since decades of research have gone
into the development of numerical methods for classical
electromagnetism, culminating in methods applicable to complicated
inhomogeneous three-dimensional geometries, it is desirable to seek
approaches for the Casimir force that exploit these developments.  We
believe that the stress-tensor approach, developed for analytical
calculations several decades ago, provides the ideal formulation for a
computational approach exploiting standard numerical techniques.

In the future, we would like to employ the stress-tensor approach to
study Casimir forces and torques in more realistic and/or more unusual
structures. The large and growing number of interesting applications
of the Casimir effect and the ongoing experimental work on
non-standard geometries~\cite{Capasso07:review} provide an
environment in which the generality and strengths of the stress-tensor
method could be exploited.  In addition, we are currently implementing
more efficient boundary-element versions of our approach in three dimensions.

We would also like to investigate related computational problems.  One
immediate possibility is to compute the net \emph{torque} on a body,
instead of the net force.  Classically, given the stress tensor
$\mathbf{T}$, one can compute the torque by integrating $\vec{r} \times
(\mathbf{T} \cdot \dS)$~\cite{Landau60}.  This has been exploited
by several authors to compute classical electromagnetic
torques~\cite{Collett03,Liu05:torque,Popescu06}.  Similarly, the
Casimir torque can be obtained by using the mean $\langle \mathbf{T}
\rangle$ from the fluctuation-dissipation theorem, just as for the net
force~\cite{lifshitz1}.  Therefore, one can compute torques by almost
the same method as above, via repeated evaluations of the classical
Green's function. Several authors have computed Casimir torques in
parallel-plate and perfect-metallic wedge
geometries~\cite{Enk95:torque, Brevik96:torque, Shao05, Razmi05,
Munday05}; other interesting structures include the Casimir
``pendulum''~\cite{Jaffe04:preprint} as well as corrugated
surfaces~\cite{Rodrigues06:torque}, although these two structures
have only been evaluated by methods with uncontrolled approximations.

\section*{Acknowledgements}

This work was supported in part by the Nanoscale Science and
Engineering Center (NSEC) under NSF contract PHY-0117795 and by a
Department of Energy (DOE) Computational Science Graduate Fellowship
under grant DE--FG02-97ER25308. D.~I.~gratefully acknowledges support
from the Netherlands Organisation for Scientific Research (NWO), under
the IRI Scheme \textit{Vernieuwingsimpuls} VIDI-680-47-209.  We are
also grateful to M. Povinelli, Federico Capasso, R. L. Jaffe, and
M. Kardar for inspiration and helpful discussions.

\section*{Appendix}

In what follows, we derive the force per unit length in
translation-invariant structures in terms of an integral over the
solutions of Bloch-periodic problems.  This is not a new idea, but an
explicit general derivation starting from the spatial integral of the
stress tensor, including the case of finite periodicity, seems
difficult to find in the literature. As given above by
\eqref{Flength}, the force per unit length can be written as:
\begin{equation}
 \frac{\vec{F}}{L} = \frac{1}{L} \int^{\infty}_0 dw \sum^{N-1}_{n=0} \iint 
 \mathbf{T}(iw; \vec{r}-n\Lambda\hat{\vec{z}}) \cdot \dS,
\label{eq:app-tensor}
\end{equation}
for a periodic structure in the $z$-direction with period $\Lambda$,
and size $L = N\Lambda$ with periodic boundary conditions (ultimately,
we will take the limit as $N\rightarrow \infty$).  As in \eqref{stress},
the stress tensor is expressed via the Green's function, and therefore
\eqref{app-tensor} can be decomposed into individual terms of the form:
\begin{equation}
  \frac{1}{L} \int^{\infty}_0 dw \sum^{N-1}_{n=0}
 \iint \mathbf{G}(iw; \vec{r}_n, \vec{r}_n) dA ,
\label{eq:app-F}
\end{equation}
where $\vec{r}_n = \vec{r} - n\Lambda\hat{\vec{z}}$ and the Green's
function is given as in \eqref{Green} by the solution of:
\begin{equation}
  \vec{G}_k(iw; \vec{x}, \vec{x'}) = \hat{O}^{-1}
 \delta^{3}(\vec{x}-\vec{x'}) \vec{\hat{e}}_k,
\label{eq:app-Gdelta}
\end{equation}
in which $\hat{O}$ denotes the linear operator $\hat{O} = \left(\nabla
\times \nabla \times + w^2 \varepsilon\right)$. In the following, we
will focus on the periodic $z$ direction and leave the $x$ and $y$
coordinates implicit for simplicity.  That is, we will write
e.g. $\delta(z - n\Lambda - z')$ instead of $\delta(\vec{r}_n -
\vec{r}')$.

At this point, we can re-express the delta function over the periodic
direction $z$ in terms of the Fourier identity:
\begin{equation}
  \delta(z-z'-n\Lambda) = \frac{1}{N} \sum^{N-1}_{\ell=0} \sum^{N-1}_{m=0}
  \delta(z-z'-\ell\Lambda -n\Lambda) e^{\frac{2\pi i}{N} \ell m},
\label{eq:app-delta-sum}
\end{equation}
Substituting this into \eqref{app-F}, we will move the $\sum_m$
outside the $\sum_n$ and consider the action of $\hat{O}^{-1}$ on the
remaining summation:
\begin{equation}
  J_{n,m} = \sum_{\ell=0}^{N-1} \delta(z-z'-\ell\Lambda -n\Lambda) e^{\frac{2\pi i}{N} \ell m} 
= J_{0,m} e^{-\frac{2\pi i}{N} m n} ,
\label{eq:app-J}
\end{equation}
where we have used the periodic boundary conditions in $L$ to realize
that $J_{n,m}$ is a cyclic shift of $J_{0,m}$ with a phase factor.  Now, we
must operate $\hat{O}^{-1}$ on $J_{0,m} \hat{\vec{e}}_k$ and evaluate at
$z' = z - n\Lambda$.  However, this corresponds to finding the field from a
Bloch-periodic current source, and such a field is also
Bloch-periodic.  Therefore:
\begin{equation}
  \left. \left( \hat{O}^{-1} J_{0,m} \hat{\vec{e}}_k \right) \right|_{z'=z-n\Lambda}
=  \left. \left( \hat{O}^{-1} J_{0,m} \hat{\vec{e}}_k \right) \right|_{z'=z} e^{\frac{2\pi i}{N} m n} ,
\label{eq:app-Green-periodic}
\end{equation}
At this point, we have completely eliminated the $n$-dependence from
the evaluation of the Green's function $\hat{O}^{-1} J_{0,m}
\hat{\vec{e}}_k$, and the phase factors from \eqref{app-J} and
\eqref{app-Green-periodic} cancel.  The remaining summation $\sum_n$
simply yields $N$, which cancels the $1/N$ factor from
\eqref{app-delta-sum}.  \Eqref{app-F} therefore becomes:
\begin{equation}
  \frac{1}{L} \int^{\infty}_0 dw \sum^{N-1}_{m=0}
 \iint \mathbf{G}(iw, m; \vec{r}_n, \vec{r}_n) dA ,
\label{eq:app-F2}
\end{equation}
where $\mathbf{G}_k(iw, m) = \hat{O}^{-1} J_{0,m} \hat{\vec{e}}_k$,
the field from a Bloch-periodic sum of delta-function sources.
Finally, we can now take the limit $N \rightarrow \infty$ by turning
$\sum_m$ into an integral:
\begin{equation}
\lim_{N\rightarrow\infty} \frac{1}{L} \sum_{m=0}^{N-1} = \frac{1}{2\pi\Lambda} \int_{-\pi/\Lambda}^{\pi/\Lambda} dk_z, 
\end{equation}
where $k_z$ is the Bloch wavevector ($k_z = 2\pi m/N$).  We therefore obtain \eqref{Freduced}.


\end{document}